\documentclass[twocolumn,tighten]{aastex6} 

\def\novak{{M31N\,2008-12a}}

\defcitealias{2014A&A...563L...9D}{DWB14}
\defcitealias{2014A&A...563L...8H}{HND14}
\defcitealias{2014ApJ...786...61T}{TBW14}
\defcitealias{2015A&A...580A..45D}{DHS15}
\defcitealias{2015A&A...580A..46H}{HND15}
\defcitealias{2015A&A...582L...8H}{HDK15}
\defcitealias{2016ApJ...833..149D}{DHB16}
\defcitealias{2016ApJ...830...40K}{KSH16}

\newcommand{\oonek}{\citetalias{2014A&A...563L...9D}}

\newcommand{\ponek}{\citetalias{2014ApJ...786...61T}}
\newcommand{\otwok}{\citetalias{2015A&A...580A..45D}}
\newcommand{\xtwok}{\citetalias{2015A&A...580A..46H}}

\newcommand{\othreek}{\citetalias{2016ApJ...833..149D}}

\begin{document} 

\received{2017 July 21}
\revised{2017 August 21}
\accepted{2017 August 22}
\published{2017 September 20}
\slugcomment{Accepted version \today, to appear in The Astrophysical Journal}

\title{No Neon, but Jets in the Remarkable Recurrent Nova \novak? ---\\ \textit{Hubble Space Telescope} Spectroscopy of the 2015 Eruption}
\shorttitle{\novak: \textit{HST} Spectroscopy of the 2015 Eruption}
\shortauthors{Darnley et al.\ 2017}

\author{
M.~J. Darnley,\altaffilmark{1}
R. Hounsell,\altaffilmark{2,3}
P. Godon,\altaffilmark{4,5}
D.~A. Perley,\altaffilmark{1}
M. Henze,\altaffilmark{6}
N.~P.~M. Kuin,\altaffilmark{7}
B.~F. Williams,\altaffilmark{8}\\
S.~C. Williams,\altaffilmark{1,9}
M.~F. Bode,\altaffilmark{1}
D.~J. Harman,\altaffilmark{1}
K. Hornoch,\altaffilmark{10}
M. Link,\altaffilmark{11}
J.-U. Ness,\altaffilmark{12}
V.~A.~R.~M. Ribeiro,\altaffilmark{13,14,15,16}\\
E.~M. Sion,\altaffilmark{4}
A.~W. Shafter,\altaffilmark{17}
M.~M. Shara\altaffilmark{18}
}

\altaffiltext{1}{Astrophysics Research Institute, Liverpool John Moores University, IC2 Liverpool Science Park, Liverpool, L3 5RF, UK}
\altaffiltext{2}{Department of Astronomy and Astrophysics, University of California, Santa Cruz, CA 95064, USA} 
\altaffiltext{3}{Astronomy Department, University of Illinois at Urbana-Champaign, 1002 W.\ Green Street, Urbana, IL 61801, USA} 
\altaffiltext{4}{Department of Astrophysics \& Planetary Science, Villanova University, 800 Lancaster Avenue, Villanova, PA 19085, USA} 
\altaffiltext{5}{Henry A.\ Rowland Department of Physics \& Astronomy, Johns Hopkins University, Baltimore, MD 21218, USA} 
\altaffiltext{6}{Institute of Space Sciences (IEEC-CSIC), Campus UAB, Carrer de Can Magrans, s/n 08193 Barcelona, Spain}
\altaffiltext{7}{Mullard Space Science Laboratory, University College London, Holmbury St.\ Mary, Dorking, Surrey RH5 6NT, UK} 
\altaffiltext{8}{Department of Astronomy, Box 351580, University of Washington, Seattle, WA 98195, USA} 
\altaffiltext{9}{Physics Department, Lancaster University, Lancaster, LA1 4YB, UK} 
\altaffiltext{10}{Astronomical Institute, Academy of Sciences, CZ-251 65 Ond\v{r}ejov, Czech Republic} 
\altaffiltext{11}{Space Telescope Science Institute, 3700 San Martin Drive, Baltimore, MD 21218, USA} 
\altaffiltext{12}{XMM-Newton Observatory SOC, European Space Astronomy Centre, Camino Bajo del Castillo s/n, Urb.\ Villafranca del Castillo, 28692 Villanueva de la Ca\~{n}ada, Madrid, Spain} 
\altaffiltext{13}{CIDMA, Departamento de F\'isica, Universidade de Aveiro, Campus de Santiago, 3810-193 Aveiro, Portugal} 
\altaffiltext{14}{Instituto de Telecomunica\c{c}\~oes, Campus de Santiago, 3810-193 Aveiro, Portugal} 
\altaffiltext{15}{Department of Physics and Astronomy, Botswana International University of Science \& Technology, Private Bag 16, Palapye, Botswana}
\altaffiltext{16}{Department of Astrophysics/IMAPP, Radboud University, P.O. Box 9010, 6500 GL Nijmegen, The Netherlands} 
\altaffiltext{17}{Department of Astronomy, San Diego State University, San Diego, CA 92182, USA} 
\altaffiltext{18}{Department of Astrophysics, American Museum of Natural History, 79th Street and Central Park West, New York, NY 10024, USA} 

\begin{abstract}
The 2008 discovery of an eruption of M31N\,2008-12a began a journey on which the true nature of this remarkable recurrent nova continues to be revealed.  M31N\,2008-12a contains a white dwarf (WD) close to the Chandrasekhar limit, accreting at a high rate from its companion, and undergoes thermonuclear eruptions which are observed yearly and may even be twice as frequent.  In this paper, we report on {\it Hubble Space Telescope} Space Telescope Imaging Spectrograph ultraviolet spectroscopy taken within days of the predicted 2015 eruption, coupled with Keck spectroscopy of the 2013 eruption.  Together, this spectroscopy permits the reddening to be constrained to $E(B-V)=0.10\pm0.03$. The UV spectroscopy reveals evidence for highly ionized, structured, and high velocity ejecta at early times.  No evidence for neon is seen in these spectra, however, but it may be that little insight can be gained regarding the composition of the white dwarf (CO versus ONe). 
\end{abstract}


\maketitle

\section{Introduction}

The eruptions of novae are among the most energetic stellar explosions in the Universe.  A nova is powered by a brief thermonuclear runaway (TNR), followed by a short-lived period of quasi-static hydrogen burning, all occurring within the surface layer of an accreting white dwarf \citep[WD; see][]{1976IAUS...73..155S}.  H-rich material is transferred from a donor star via an accretion disk around the WD, possibly with some contribution from magnetic accretion.   The TNR powers an  ejection of the accreted material, with a rapidly expanding pseudo-photosphere initially increasing the visible luminosity of the system by up to eight orders of magnitude; the brightest novae can reach $M_V<-10$ \citep{2009ApJ...690.1148S}.  \citet{2008clno.book.....B}, \citet{2010AN....331..160B}, and \citet{2014ASPC..490.....W} present recent comprehensive reviews of these phenomena.

The majority of known nova systems, classical novae (CNe), have only been observed in eruption once.  Recurrent novae (RNe) have been observed in eruption at least twice, most numerous times \cite[see][for a summary]{2010ApJS..187..275S}.  Short recurrence periods are driven by a combination of a high WD mass and a high accretion rate.  Such high accretion rates are typically achieved via Roche-lobe overflow of a sub-giant donor, or from the stellar wind of a red giant donor \citep[see][]{2012ApJ...746...61D}.  Observed recurrence times ($P_\mathrm{rec}$) range from 1\,year \citep{2014A&A...563L...9D} up to 98\,years \citep{2009AJ....138.1230P}.  Both ends of this observed range are probably constrained only by selection effects.  Until recently, the contribution of RN systems ($1\le P_\mathrm{rec}\le100$\,yrs) to the nova population was believed to be only at the level of a few percent.  Studies by \citet{2014ApJ...788..164P}, \citet{2015ApJS..216...34S}, and \citet{2016ApJ...817..143W} independently indicated that RNe, at least in late-type galaxies, may account for around 1/3 of all nova eruptions.

Over 1000 nova candidates have been discovered in M\,31 \citep[and see their on-line catalog\footnote{\url{http://www.mpe.mpg.de/~m31novae/opt/m31/index.php}}]{2007A&A...465..375P,2010AN....331..187P}, with a large proportion of recent candidates spectroscopically confirmed \citep{2011ApJ...734...12S}.  Such studies of M\,31 led to the discovery of a new class of rapid recurrent novae (RRNe) all with $P_\mathrm{rec}<10$\,years.  Currently there are eight members of this class, seven in M\,31 and one in the Large Magellanic Cloud (LMC); the prototype RRN is \novak.

At the time of writing, \novak\ has been detected in eruption nine times in nine consecutive years \citep[2008--2016;][]{2014A&A...563L...9D,2015A&A...580A..45D,2016ApJ...833..149D,2016ATel.9848....1I}\footnote{Also see \citet{2016A&A...593C...3D}}, with three additional earlier detections recovered from archival X-ray observations \citep[see Table~\ref{eruption_history}]{2014A&A...563L...8H,2014ApJ...786...61T}, for a detailed summary of all detected eruptions, see Table~1 within \citet[hereafter \othreek]{2016ApJ...833..149D}.  The eight 2008--2015 eruptions point to a recurrence period of $P_\mathrm{rec}=347\pm10$\,days, but the addition of the three earlier eruptions may require an even shorter period of $P_\mathrm{rec}=174\pm10$ \citep{2015A&A...582L...8H,2016ApJ...833..149D}.

\begin{table}
\caption{List of all observed eruptions of \novak.\label{eruption_history}}
\begin{center}
\begin{tabular}{lll}
\hline
\hline
Eruption date\tablenotemark{a} & Days since & References\\
(UT) & last eruption\tablenotemark{b} \\
\hline
(1992 Jan.\ 28) & \nodata & 1, 2 \\
(1993 Jan.\ 03) & 341 & 1, 2 \\
(2001 Aug.\ 27) & \nodata & 2, 3 \\
2008 Dec.\ 25 & \nodata & 4 \\
2009 Dec.\ 02 & 342 & 5 \\
2010 Nov.\ 19 & 352 & 2 \\
2011 Oct.\ 22.5 & 337.5 & 5, 6--8 \\
2012 Oct.\ 18.7 & 362.2 & 8--11 \\
2013 Nov.\ $26.95\pm0.25$ & 403.5 & 5, 8, 11--14 \\
2014 Oct.\ $02.69\pm0.21$ & $309.8\pm0.7$ & 8, 15 \\
2015 Aug.\ $28.28\pm0.12$ & $329.6\pm0.3$ &14, 16--18\\
2016 Dec.\ 12.32 & 471.72 & 19, 20\\
\hline
\end{tabular}
\end{center}
\catcode`\&=12
\tablenotetext{a}{Estimated date of eruption, those in parentheses are extrapolated from X-ray data \protect \citep[see][]{2015A&A...582L...8H}.}
\tablenotetext{b}{Inter-eruption time quoted for consecutive detections in consecutive years, assuming of $P_\mathrm{rec}\simeq1$\,year.}
\tablecomments{Updated version of Table~1 from \protect \citet{2014ApJ...786...61T}.}
\tablerefs{(1)~\citet{1995ApJ...445L.125W}, (2)~\citet{2015A&A...582L...8H}, (3)~\citet{2004ApJ...609..735W}, (4)~\citet{2008Nis}, (5)~\citet{2014ApJ...786...61T}, (6)~\citet{2011Kor}, (7)~\citet{2011ATel.3725....1B}, (8)~\citet{2015A&A...580A..45D}, (9)~\citet{2012Nis}, (10)~\citet{2012ATel.4503....1S}, (11)~\citet{2014A&A...563L...8H}, (12)~\citet{2013ATel.5607....1T}, (13)~\citet{2014A&A...563L...9D}, (14)~\citet{2016ApJ...833..149D}, (15)~\citet{2015A&A...580A..46H}, (16)~\citet{2015ATel.7964....1D}, (17)~\citet{2015ATel.7965....1D}, (18)~\citet{2015ATel.7984....1H}, (19)~\citet{2016ATel.9848....1I}, (20)~\citet{12a2016}.}
\end{table}

The short recurrence period and rapid, under-luminous, eruptions of \novak\ point to a very high mass WD and large mass accretion rate.  Based on the 2013 eruption alone, \citet[hereafter \ponek]{2014ApJ...786...61T} constrained the WD mass to be $>1.3\,M_\odot$.  Whereas, using data from multiple eruptions, \citet{2015ApJ...808...52K} indicated that $M_\mathrm{WD}=1.38\,M_\odot$.  The \ponek\ and \citet{2015ApJ...808...52K} analyses suggested WD accretion rates of $\dot{M}_\mathrm{acc}>1.7\times10^{-7}\,M_\odot\,\mathrm{yr}^{-1}$ and $\dot{M}_\mathrm{acc}=1.6\times10^{-7}\,M_\odot\,\mathrm{yr}^{-1}$, respectively.  The \citet{2015ApJ...808...52K} models also predicted a low ejected mass of $\sim6\times10^{-8}\,M_\odot$, leading to a proposed mass accumulation efficiency of 63\%.  \citet[hereafter \xtwok]{2015A&A...580A..46H} determined an ejected H mass of $M_\mathrm{e,H}=\left(2.6\pm0.4\right)\times10^{-8}\,M_\odot$, but we note that this assumes a spherical ejecta geomoetry.  \othreek\ noted that eruptions of \novak\ showed some of the most rapid developments of any known nova.  The time taken for the nova to decay by two magnitudes from peak ($t_2$) is only $1.65\pm0.04$\,days, and the system is the only known nova with $t_3<3$\,days.  Spectroscopically, the eruptions of \novak\ briefly exhibit the highest expansion velocities yet observed from a nova eruption ($v_\mathrm{ej}\simeq13,000$\,km\,s$^{-1}$; \othreek) and provided tentative evidence for collimated polar outflows.  The spectral line evolution in the optical is consistent with the ejecta interacting with and shocking pre-existing circumbinary material \citep[hereafter \otwok]{2015A&A...580A..45D}, proposed to be a stellar wind from a red giant donor (\othreek).  The high mass WD, which is seemingly growing, suggests that \novak\ is a prime pre-explosion supernova type Ia candidate.  A brief review of M31N\,2008-12a, up to and including the 2015 eruption, is given in \citet{2017ASPC..509..515D}.

In this paper, we present the results of a {\it Hubble Space Telescope} (HST) spectroscopic program to study the predicted 2015 eruption of \novak\ and an updated analysis of archival Keck observations of the system.  In Section~\ref{sec:observations} we describe the observations on which this work is based.  Section~\ref{sec:extinction} addresses the interstellar extinction toward \novak.  In Sections~\ref{sec:spec} we present the spectroscopic data.  Finally in Sections~\ref{sec:disc} and \ref{sec:conc} we discuss our findings and present our conclusions.

During the preparation of this manuscript the 2016 eruption of \novak\ was discovered \citep{2016ATel.9848....1I}.  Results from observations of that eruption will be presented in \citet{12a2016}.

\section{Observations of the 2015 Eruption}\label{sec:observations}

\subsection{{\textit Hubble Space Telescope} UV Spectroscopy}

We were awarded 20 orbits of Cycle\,23 {\it HST} disruptive target of opportunity (DToO) time to obtain both early-time UV spectroscopic observations and late-time imaging data of the 2015 eruption of \novak\ (proposal ID: 14125), the late-time photometry will be reported in \citet{2017arXiv170910145D}.  The 2015 eruption of \novak\ was discovered on 2015 Aug.\ 28.425\,UT by the Las Cumbres Observatory 2\,m telescope\footnote{Formerly known as the Faulkes Telescope North} on Hawai'i \citep{2015ATel.7964....1D}, the DToO was triggered at Aug.\ 28.65.   The {\it HST} spectroscopic observations were conducted between 2015 Aug.\ 31 and Sep.\ 2, a log of these observations is provided in Table~\ref{obs_log}.

\begin{table*}
\caption{Log of observations of the eruptions of \novak\ referred to in this Paper.\label{obs_log}}
\begin{center}
\begin{tabular}{lllllllll}
\hline
\hline
Eruption & Facility & Instrument & {\it HST}  & Date & Start & End & Orbits & Exposure\\
     &   &    & Visit              & (midpoint) & \multicolumn{2}{c}{$t-t_0$ (days)} & & time (ks)\\
\hline
2013 & Keck I & LRIS & \nodata & 2013 Dec 02.23 & \phn5.27 & \phn5.29 & \nodata & \phn1.3 \\
\hline
2015 & {\it HST} & STIS/FUV MAMA & 1 & 2015 Aug 31.60 & \phn3.20 & \phn3.44 & 4 & 10.2\\
2015 & {\it HST} & STIS/NUV MAMA & 2 & 2015 Sep 01.59 & \phn4.26 & \phn4.37& 2 & \phn4.8\\
2015 & {\it HST} & STIS/NUV MAMA & 3 & 2015 Sep 02.59 & \phn5.25 & \phn5.36 & 2 & \phn4.8\\
\hline
\end{tabular}
\end{center}
\end{table*}

Early eruption UV spectroscopy was obtained using the Space Telescope Imaging Spectrograph (STIS) across both the NUV and FUV regimes. These observations were taken using the G230L (1570--3180\,\AA) and G140L (1150--1737\,\AA) gratings, with a $52^{\prime\prime}\times0^{\prime\prime}\!\!.2$ slit configuration, and the MAMA detectors.  Although offset in time, when combined these gratings provide an effective wavelength range of 1150--3180\,\AA\ with a small overlap between 1650--1709\,\AA.  STIS target acquisition utilized a nearby bright optical stellar source (HSTID:\,NBW9014542) and the recommended F28X50LP aperture.   

The maximum number of consecutive orbits in which the MAMA detectors can be employed is five. As a consequence of this limitation the FUV and NUV observations were split into two separate visits $\sim1$\,day apart, with the FUV observations occurring first due to their lower throughput. At the time of trigger an additional  scheduling limitation occurred, resulting in the splitting of the NUV visit over two consecutive days.

In order to remove detector defects and hot pixels a four point STIS-ALONG-SLIT dither pattern was initiated for the four orbit FUV observations, with a two point dither for each of the two orbit NUV visits. Each dither was applied with a seven pixel point spacing (FUV $0^{\prime\prime}\!\!.168$, NUV $0^{\prime\prime}\!\!.175$) to allow for improved spatial resolution.

Raw STIS data were processed via the STScI {\tt calstis} pipeline to produce a variety of calibrated data files. Processes implemented by the pipeline include: 2D image reduction (i.e, subtraction of the over-scan region and bias subtraction); cosmic ray rejection; dark subtraction and application of flat fields; obtaining zero-point shifts in the spectral and spatial directions; conducting  flux/wavelength calibrations; and summation of spectral exposures \citep[see][]{2012stii.book.....H}. Note that raw MAMA data is of the $2048 \times 2048$ pixel format, but once calibrated it is binned via the pipeline to $1024\times1024$ pixels. 

\subsection{Keck spectroscopy of the 2013 eruption}

To supplement the {\it HST} UV spectra, and provide context over a larger wavelength range, we will also refer to optical spectra taken at similar epochs during other eruptions of \novak.  During the 2015 eruption, the Liverpool Telescope obtained optical spectra at similar epochs to the {\it HST} spectra ($t=3.84$ and 4.91\,days; \othreek), however these spectra had very low signal-to-noise (S/N), and are not overly useful for this purpose.

\othreek\ demonstrated that the 2013, 2014, and 2015 eruptions were essentially identical, at all times throughout the evolution, from the optical to the X-ray.  Therefore, for comparative purposes we turn to the Keck spectrum of the 2013 eruption taken at $t=5.28$\,days post-eruption.  This spectrum was originally published by \ponek\ however, here we have re-analyzed that spectrum in the context of what has been learned from subsequent eruptions of \novak.  The spectrum was obtained using the Low Resolution Imaging Spectrometer \citep[LRIS;][]{1995PASP..107..375O,1998SPIE.3355...81M,2010SPIE.7735E..0RR} instrument mounted at the Cassegrain focus of the Keck I telescope on Mauna Kea, Hawai'i.  The spectrum was acquired using a standard low-resolution configuration using the 400/3000 grism (blue camera) and 400/8500 grating (red camera) to provide continuous coverage from the atmospheric cutoff to approximately 10300\,\AA. 
 
\section{Extinction}\label{sec:extinction} 

A good understanding of the scale of the dust extinction and related gas absorbing column between \novak\ and ourselves is crucial for furthering our understanding of this system.  In particular, the interpretation of the UV observations in this work, the UV and X-ray observations reported by \citet{2014A&A...563L...8H}, \otwok, \xtwok, and \othreek, and the theoretical work by \citet{2014ApJ...793..136K,2015ApJ...808...52K,2016ApJ...830...40K,2017ApJ...838..153K} are especially sensitive to the adopted value of the extinction.

In previous work, we assumed that the line of sight extinction is comprised of two components, the foreground Galactic reddening \citep[assumed to be $E^\mathrm{FG}_{B-V}=0.10$;][]{1992ApJS...79...77S}, and an unknown component internal to M\,31.  In \citet[hereafter \oonek]{2014A&A...563L...9D} we employed the $6^{\prime\prime}$ ($\sim100$\,pc) resolution M\,31 dust maps of \citet{2009A&A...507..283M} to estimate the maximal internal reddening at the position of \novak\ to be $E^\mathrm{M31}_{B-V}=0.16$.  In that, and subsequent papers, we therefore adopted a total reddening in the range $0.10\le E_{B-V}\le0.26$.  \ponek\ utilized optical spectra of the 2013 eruption to estimate the total reddening towards \novak\ to be $E_{B-V}=0.24\pm0.09$.  This was calculated using the H$\alpha$/H$\beta$ ratio and assuming case B recombination.  However, we note that such an approach has often been problematic with novae due to, e.g., self-absorption within the ejecta.

Model spectrum fitting to the combined 2013, 2014, and 2015 X-ray light-curves of \novak\ can be used to constrain the line of sight absorbing neutral H column.  This fitting requires a total column (foreground plus M\,31) of $N_{\mathrm{H}}=0.3^{+0.4}_{-0.3}\times10^{21}$\,cm$^{-2}$ (\othreek).  Such a column equates to a total reddening of $E_{B-V}=0.05^{+0.07}_{-0.05}$, when using the conversion factor given by \citet{2009MNRAS.400.2050G}.

The Panchromatic Hubble Andromeda Treasury \citep[PHAT;][]{2012ApJS..200...18D} survey has recently published even higher resolution (25\,pc) dust extinction maps of M\,31 \citep{2015ApJ...814....3D}.  This work finds that the fraction of M\,31 stars, at the position of \novak, suffering from M\,31 internal reddening is 0.33.  Those stars, which are behind the M\,31 dust, experience extinction whose strength follows a log--normal distribution with median $\widetilde{A_{V}}=0.984$ and dimensionless $\sigma=0.3$ (J.~J.\ Dalcanton, priv.\ comm.).  Using the relations within \citet{2015ApJ...814....3D}, these relate to a mean M\,31 internal extinction $\bar{A_{V}}=1.03$, $\sigma_{\mathrm{SD}}=0.32$, or $E_{B-V}=0.332$, $\sigma_{\mathrm{SD}}=0.102$ -- plus the foreground reddening.  So the PHAT extinction maps predict that there is a 67\% probability that the reddening towards \novak\ is $E^\mathrm{FG}_{B-V}\left(=0.10\right)$, or a 33\% chance that it is $E^\mathrm{FG}_{B-V}+0.3\pm0.1\left(=0.4\pm0.1\right)$.

\begin{figure*}
\centering\includegraphics[width=\textwidth]{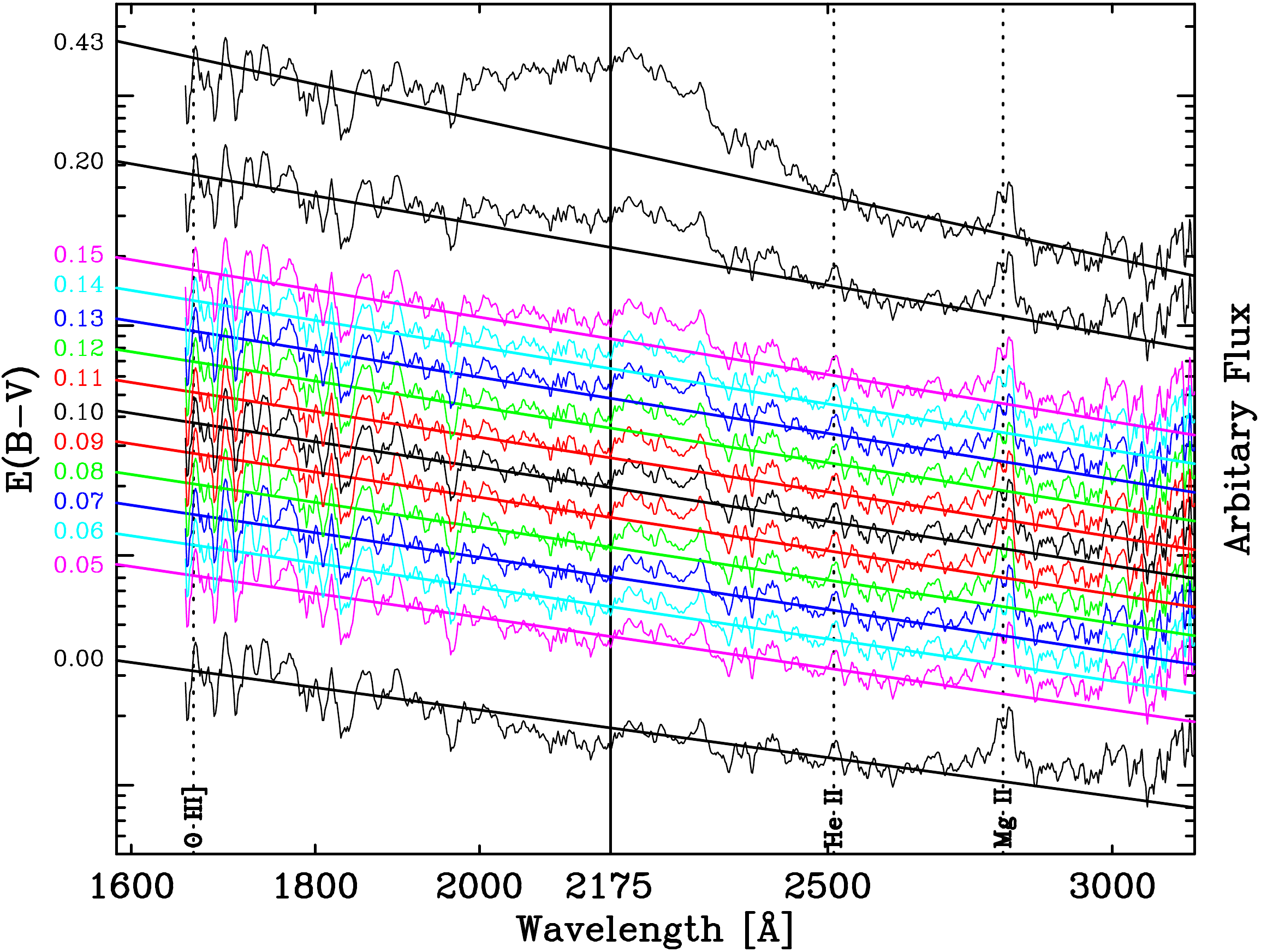}
\caption{{\it HST} STIS NUV spectrum of \novak\ de-reddened assuming different values of $E_{B-V}$ as indicated on the left-hand axis, each spectrum has been smoothed by a `boxcar' function of width 7 pixels.  A vertical line has been plotted to indicate the location of the 2175\,\AA\ spectral feature due to interstellar reddening.  The bottom spectrum (black line) shows the observed spectrum with no de-reddening, the interstellar absorption feature can be seen as a $\sim200$\,\AA\ half-width feature blueward of 2175\,\AA, redward the feature is confused by emission lines.  The de-reddened spectra are shown in color for clarity.  As the value of $E_{B-V}$ is increased towards $\sim0.1$ the broad absorption feature disappears in the de-reddened spectra, with further increases in assumed extinction the feature turns into emission.  The straight lines indicate power-law fits to each spectrum (excluding the 2175\,\AA\ feature and any prominent lines), and are included to aid comparison of the spectra.  From this plot, we adopt $E_{B-V}=0.10\pm0.03$ (the central black spectrum), consistent with foreground Galactic extinction and no extinction internal to M\,31.\label{fig:ext}}
\end{figure*}

Here we utilize the {\it HST} STIS NUV spectra of the 2015 eruption of \novak\ to constrain the reddening towards the system.  The faintness of \novak, at the distance of M\,31, and the low resolution of these STIS spectra remove any realistic opportunity to determine the reddening via high-resolution spectra of interstellar lines (see Section~\ref{ism_absorb}).  As such, we will employ the 2175\,\AA\ feature \citep{1965ApJ...142.1683S} to determine the reddening using a modified version of the procedure outlined by \citet{1987A&AS...71..339V}.  This technique entails gradually increasing the de-reddening applied to a NUV spectrum until the 2175\,\AA\ `bump' effectively disappears, with the `best' value generally determined by visual inspection.  To de-redden  the spectra, we employed the empirical extinction function of \citet{1989ApJ...345..245C} and assumed $R_V=3.1$.  A series of stacked and smoothed NUV spectra of \novak\  are shown in Figure~\ref{fig:ext} on a log--log plot, the amount of de-reddening applied is indicated on the y-axis.  For each value of $E_{B-V}$ a best-fit\footnote{All spectral fitting discussed in this paper employs a nonlinear least-squares Levenberg--\citet{doi:10.1137/0111030} algorithm, with errors on the dependant and independent variables evaluated using Orear's effective variance method \citep{doi:10.1119/1.12972}.} power law (fitted to the spectra blueward of Mg\,{\sc ii} (2800\,\AA) and excluding the 2175\,\AA\ feature) is also plotted.  The 2175\,\AA\ feature is clearly visible in the $E_{B-V}=0$ spectrum as a broad absorption beginning at $\sim2000$\,\AA, with half-width of $\sim200$\,\AA.  However, the red side of the feature may be contaminated with emission lines and is ignored for this analysis.  

The effect of increasing de-reddening \citep[up to the maximum predicted by][]{2015ApJ...814....3D} is illustrated in Figure~\ref{fig:ext}, where the 2175\,\AA\ feature evolves from a shallow dip to a broad peak.  Our adapted methodology directly compares the consistency between pairs of the best-fitting power-laws produced when either excluding or including the 2175\,\AA\ feature in the fit.  We conclude that the reddening towards \novak\ must be $E_{B-V}=0.10\pm0.03$ (1$\sigma$ uncertainties).  This value is consistent with all the extinction being in the foreground (Galactic) and none internal to M\,31 (i.e.\ \novak\ falls within the `67\%').  Such a level of extinction is also consistent with the X-ray light-curve fitting, and within $2\sigma$ of the value derived from the optical spectral analysis by \ponek.  We therefore adopt a reddening of $E_{B-V}=0.10$ throughout.  Based on this work, this value was also adopted by \othreek.

\section{Eruption spectroscopy}\label{sec:spec}

\begin{figure*}
\begin{center}
\includegraphics[width=0.97\columnwidth]{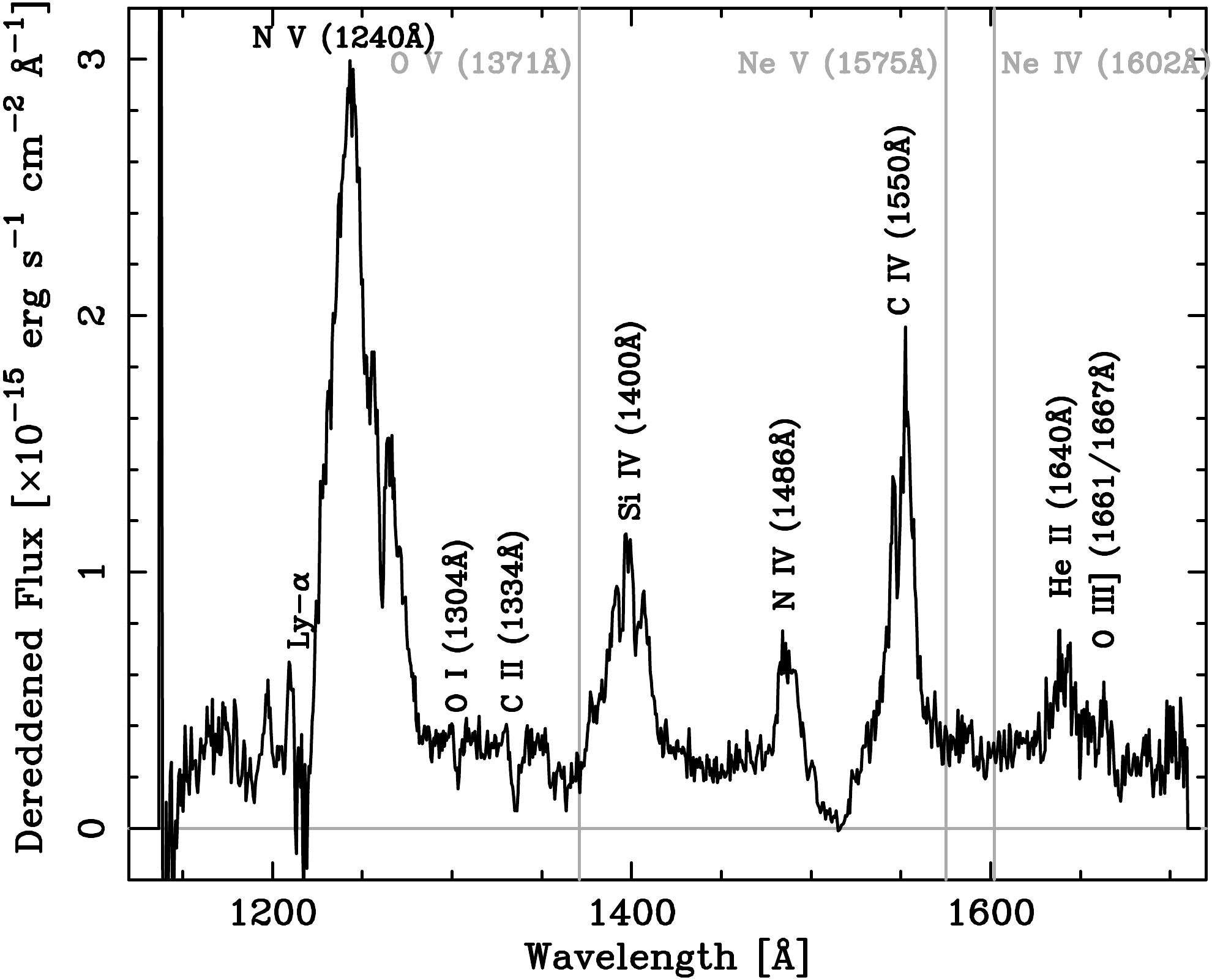}\hfill
\includegraphics[width=0.97\columnwidth]{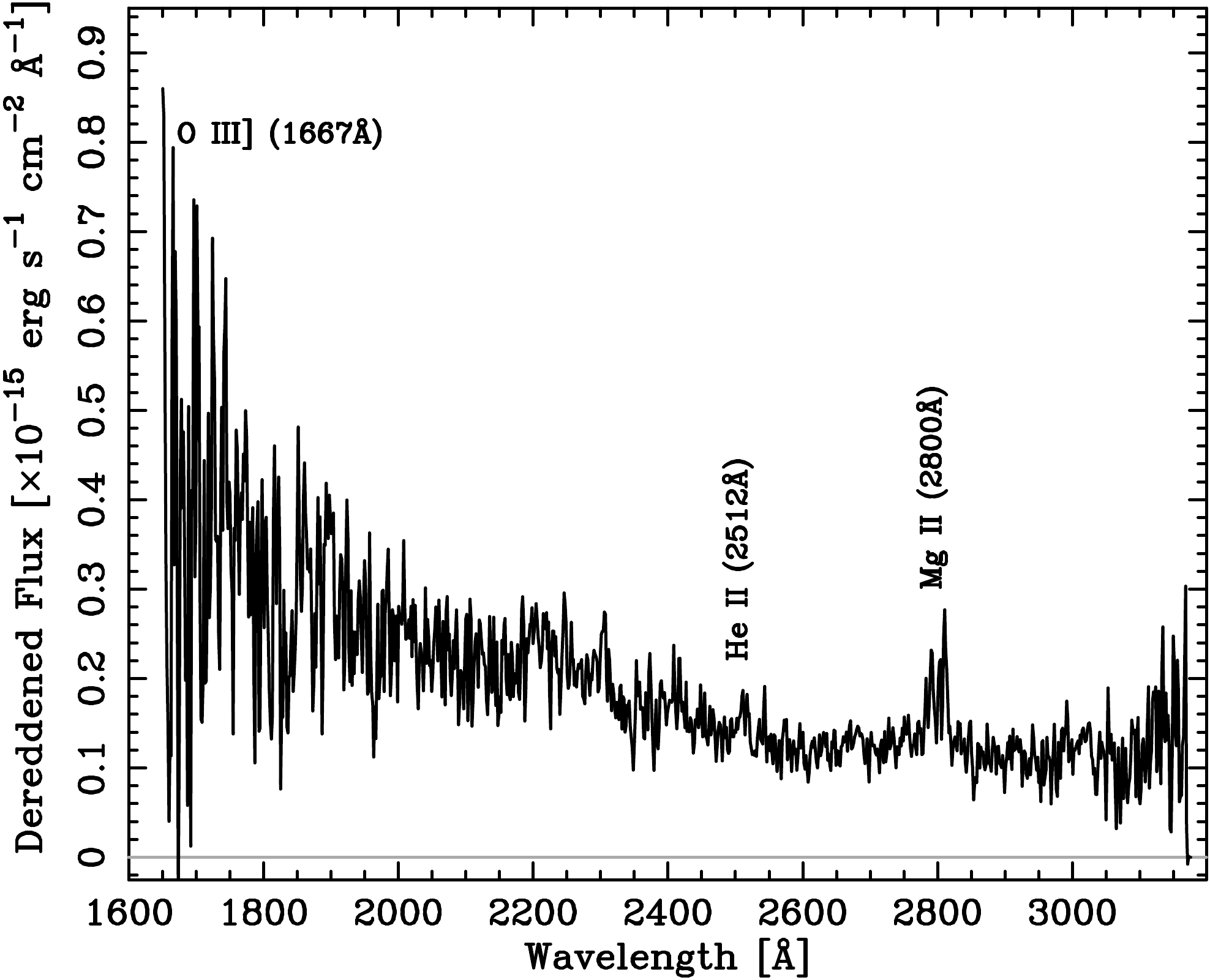}
\vfill
\includegraphics[width=0.97\textwidth]{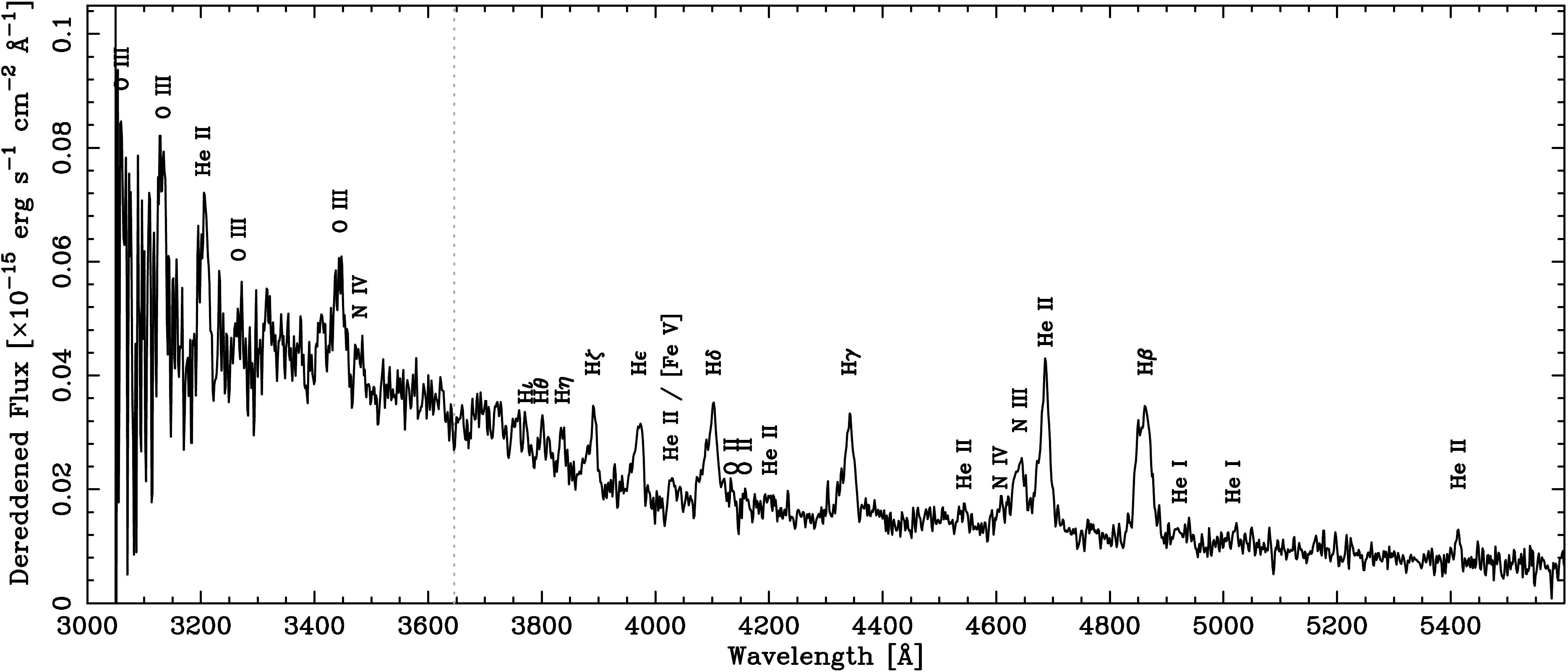}\\
\vfill
\includegraphics[width=0.97\textwidth]{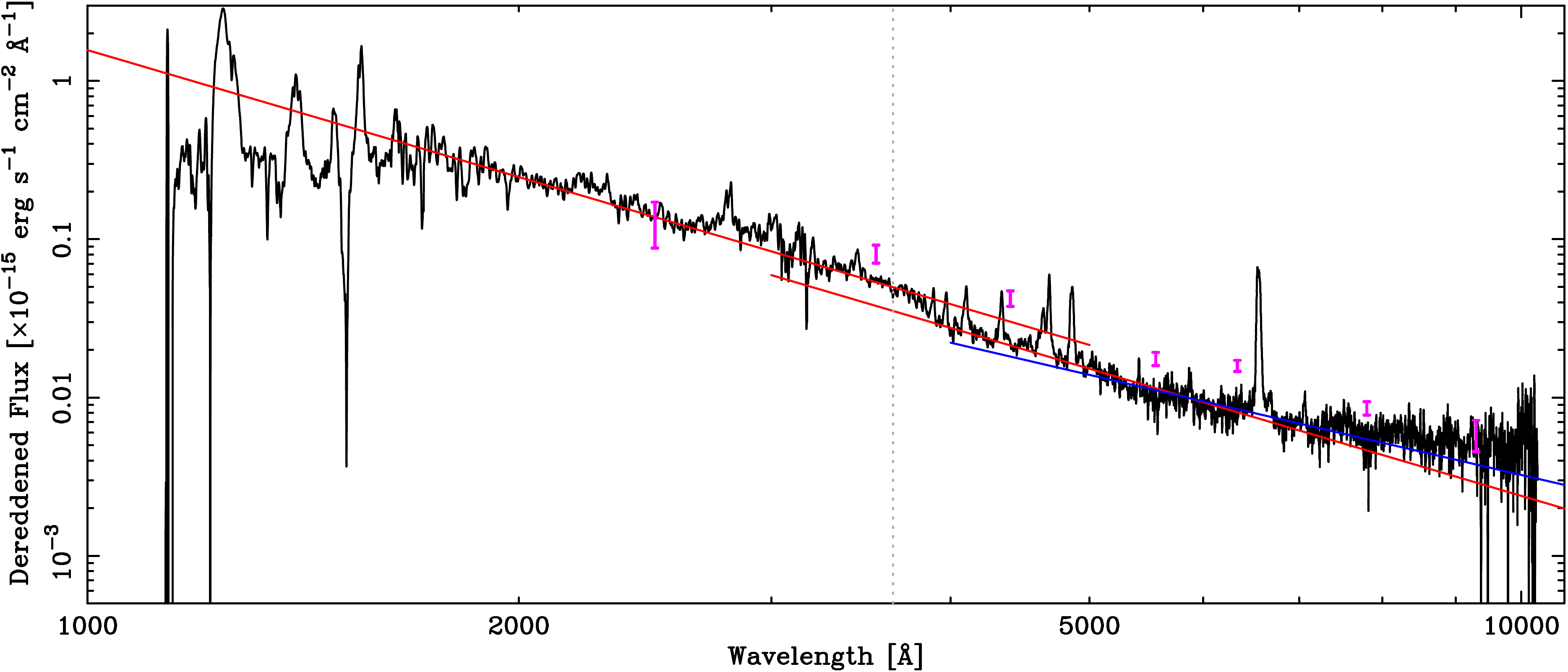}\\
\end{center}
\caption{{\bf Top:} {\it Hubble Space Telescope} STIS spectra of the 2015 eruption of \novak, left: STIS FUV MAMA G140L spectrum, right: STIS NUV MAMA G230L, both spectra have been corrected for interstellar extinction ($E_{B-V}=0.1$).  {\bf Middle:} Keck spectrum of the 2013 eruption taken at a similar epoch.  The vertical dotted line in this and the bottom plot indicate the Balmer limit (3645\,\AA). {\bf Bottom:} Combined FUV, NUV, Keck spectrum showing fits to the continuum and broadband SED (magenta points), see text for further details.\label{main_spec}}
\end{figure*}

No significant variation is observed between the 4 {\it HST}/STIS FUV orbits, so these have been combined into a single spectrum taken $3.32\pm0.12$\,days after the 2015 eruption of \novak.  This epoch is approximately coincident with the $t_3$ timescale in the optical ($2.75\pm0.28$\,days; \othreek) but significantly before $t_3$ in the NUV ($5.65^{+0.22}_{-0.40}$\,days).  The S/N of the STIS NUV spectra taken one and two days later is low, so these have been combined into a single epoch spectrum with $t=4.81\pm0.55$\,days, roughly consistent with the NUV $t_3$ and the `turn on' of the super-soft source (SSS) X-ray emission ($t=5.6\pm0.7$\,days; \othreek).  These stacked and subsequently de-reddened spectra ($E_{B-V}=0.1$; see Section~\ref{sec:extinction}) are presented in the top panel of Figure~\ref{main_spec}, prominent lines are labeled with their most probable identification. The presented spectra have not been corrected for the relative velocity of \novak.  M31 has a radial velocity of $-300$\,km\,s$^{-1}$ \citep{2012AJ....144....4M}, however the north-west side of that galaxy is receding with a rotational velocity of $\sim226$\,km\,s$^{-1}$ \citep{2006ApJ...641L.109C}, there is no indication from these spectra that the peculiar velocity of \novak\ is signficant.  Therefore, the net velocity of \novak\ relative to the Milky Way is expected to be small, and smaller than the spectral resolution.

The FUV spectrum shows a number of bright and broad emission lines.  The N\,{\sc v} (1239/1243\,\AA) and C\,{\sc iv} (1548/1551\,\AA) resonant doublets, and the Si\,{\sc iv}/O\,{\sc iv} blend at $\sim1400$\,\AA\,  dominate the FUV spectrum, showing strong emission peaks and evidence for deep P\,Cygni absorption components.  Lines of N\,{\sc iv}] (1486\,\AA), He\,{\sc ii} (1640\,\AA), and O\,{\sc iii}] (1661/1667\,\AA) are also present in emission.  Many of the bright emission lines appear to contain narrower absorption features, which might be directly related to the ejecta morphology or in some cases to interstellar absorption.  Additional absorption features are seen at $\sim$1304 and $\sim$1334\,\AA, probably corresponding to O\,{\sc i} and C\,{\sc ii}, respectively.  A Lyman\,$\alpha$ feature is present, but it is blended with the N\,{\sc v} P\,Cygni absorption.  Unidentified emission lines may also be present at $\sim$1197 and $\sim$1211\,\AA.

In the combined NUV spectrum only the Mg\,{\sc ii} (2796/2803\,\AA) resonant doublet is clearly identifiable.  Emission lines of He\,{\sc ii} (2511\,\AA) and O\,{\sc iii}] (1667\,\AA) are tentatively detected.  Currently unidentified emission lines may also appear at 1818, 1857 (possibly Al\,{\sc iii} 1860\,\AA), 1897, and 2304\,\AA, with many of these appearing double-peaked.  However, the S/N in the NUV decreases significantly blueward of $\sim1950$\,\AA. There is no evidence for any P\,Cygni absorption features in the NUV.

UV spectra of nova eruptions evolve through three clearly identifiable stages; early epochs where Fe absorption lines dominate, the `Fe curtain' phase; a transition (or pre-nebular) phase; before evolution into a nebular spectrum \citep[see][for a recent review]{2012BASI...40..185S}.  The FUV and NUV spectra of the 2015 eruption of \novak\ clearly indicate a nova eruption in the transition phase, well past the Fe curtain. 

The 2013 Keck spectrum, taken at a similar eruption epoch to the {\it HST} NUV spectrum, shows numerous strong and weak emission lines, dominated by H\,{\sc i}, He\,{\sc i}, He\,{\sc ii}, O\,{\sc iii} and various N lines.  As reported by \otwok, and particularly \othreek, there is no evidence for low-ionization Fe or O\,{\sc i} emission lines -- key signatures of the Fe\,{\sc ii} spectroscopic class -- observed in the Keck or {\it HST} spectra.

\begin{figure*}
\includegraphics[width=\columnwidth]{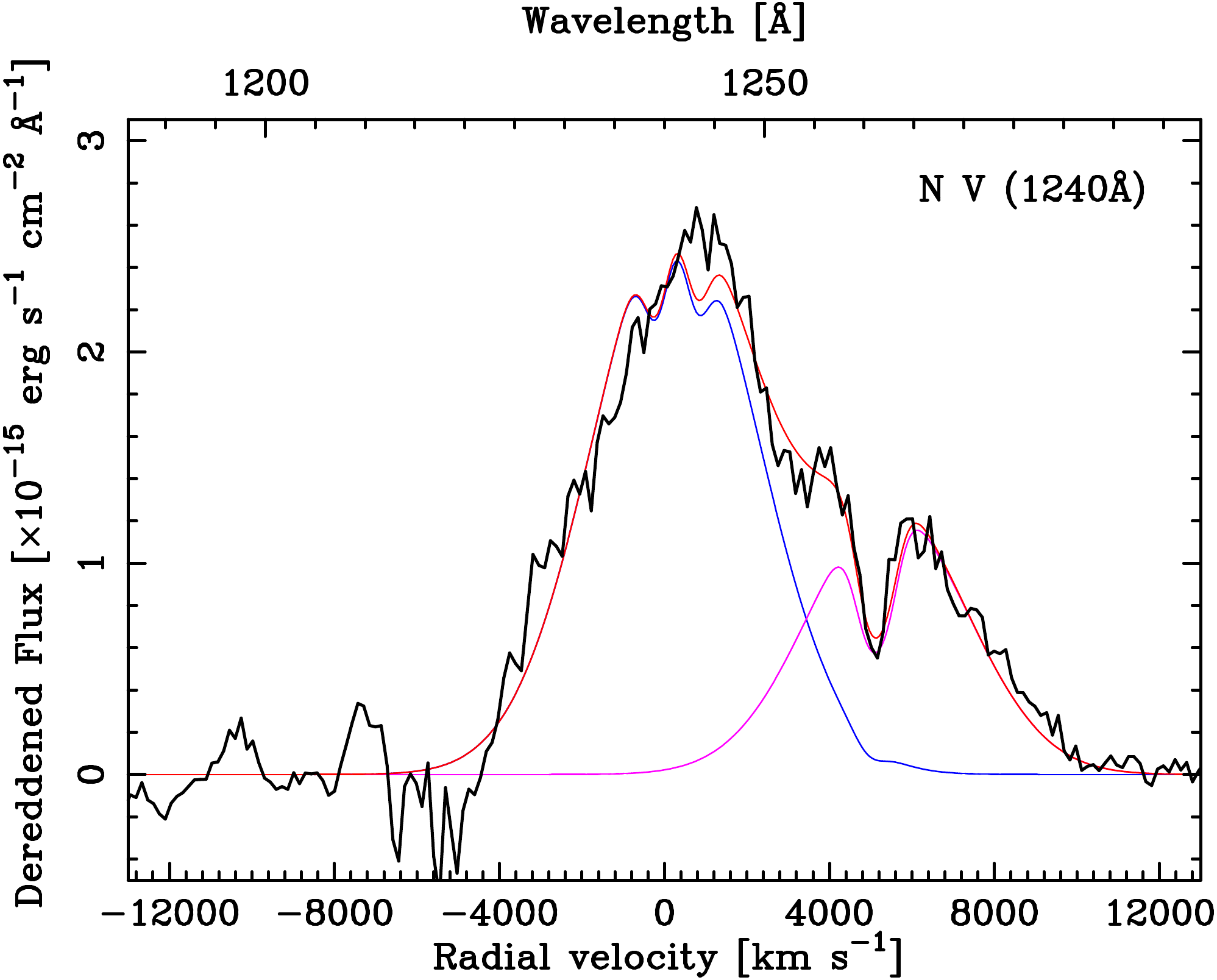}\hfill
\includegraphics[width=\columnwidth]{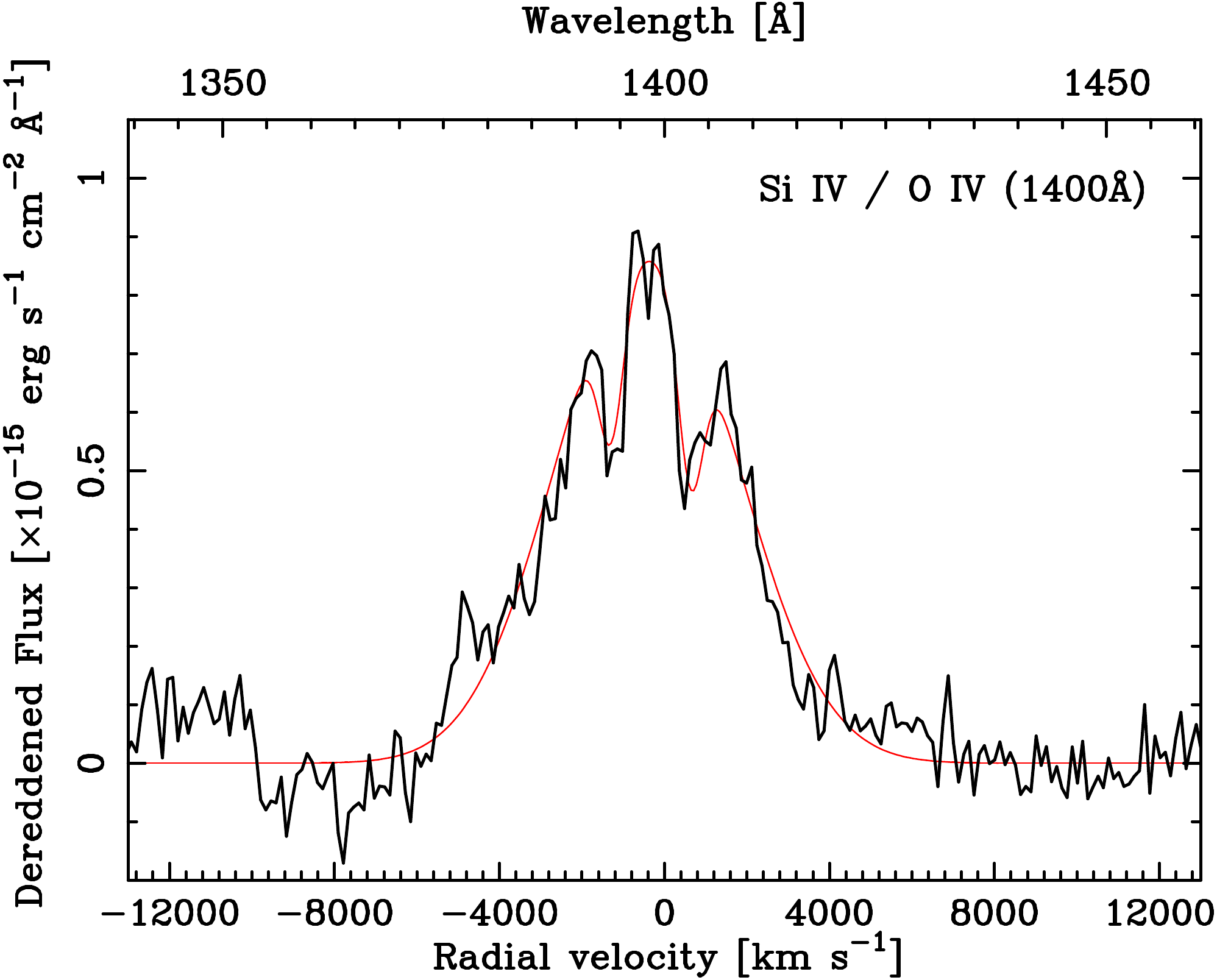}\\
\vfill
\includegraphics[width=\columnwidth]{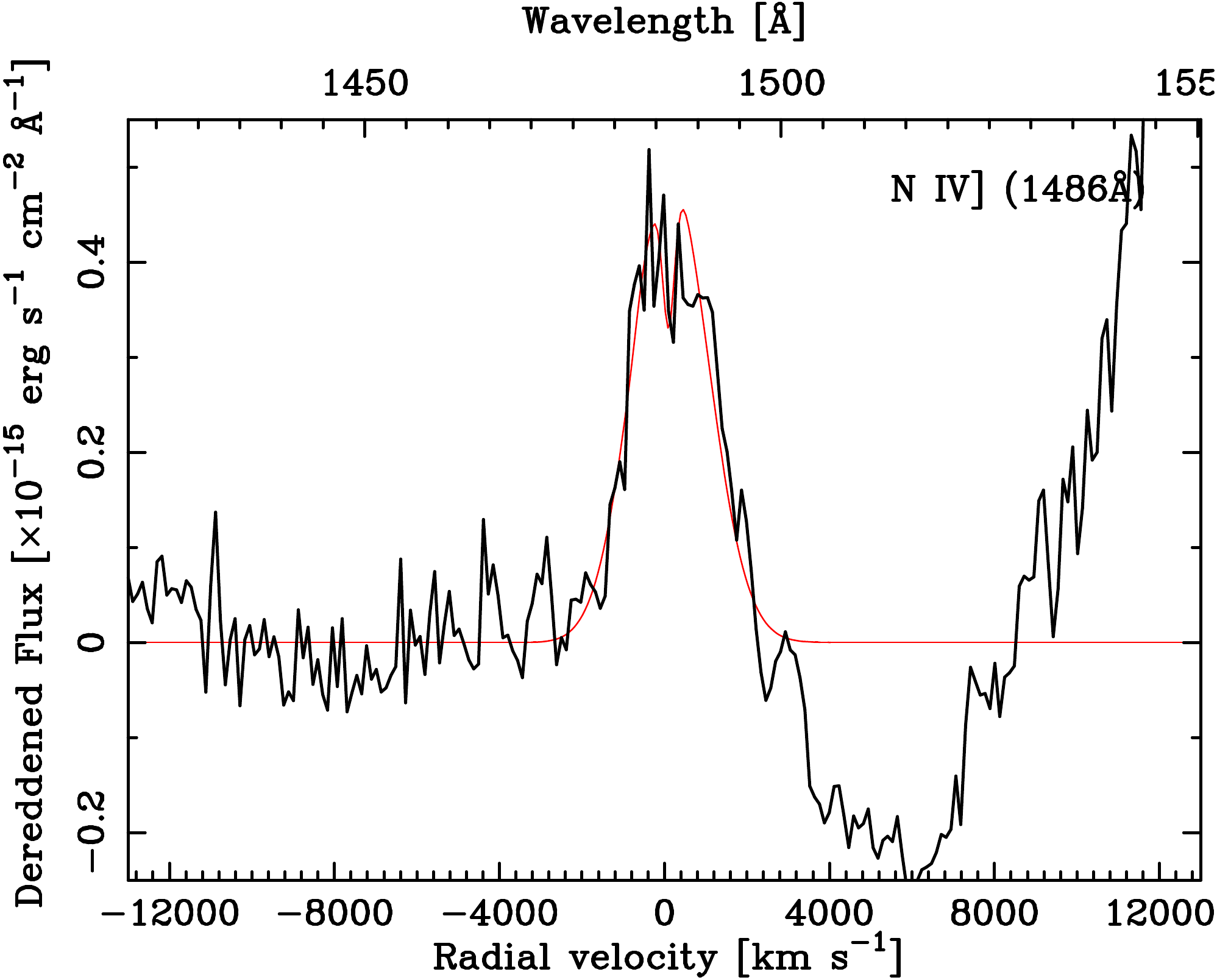}\hfill
\includegraphics[width=\columnwidth]{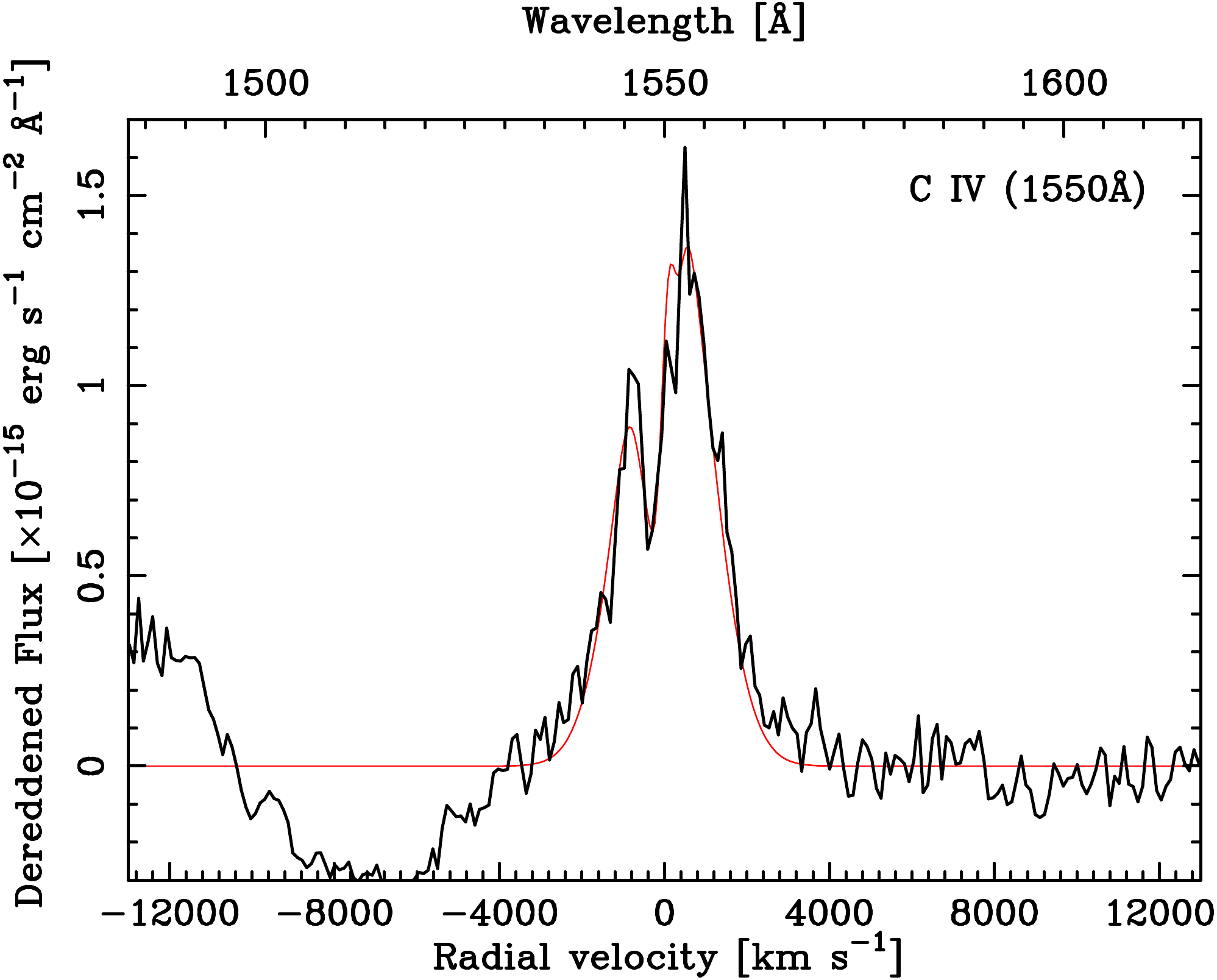}\\
\vfill
\includegraphics[width=\columnwidth]{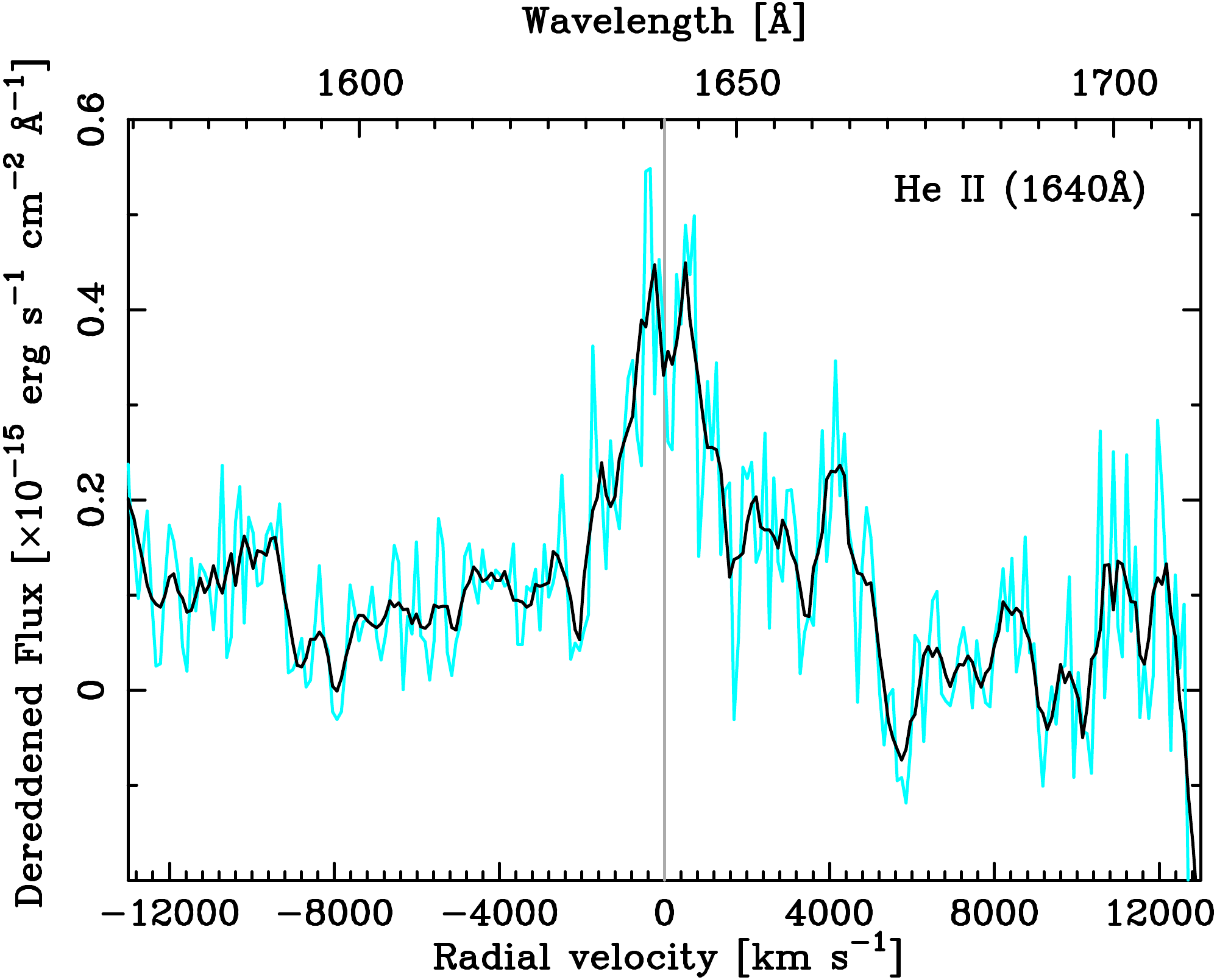}\hfill
\includegraphics[width=\columnwidth]{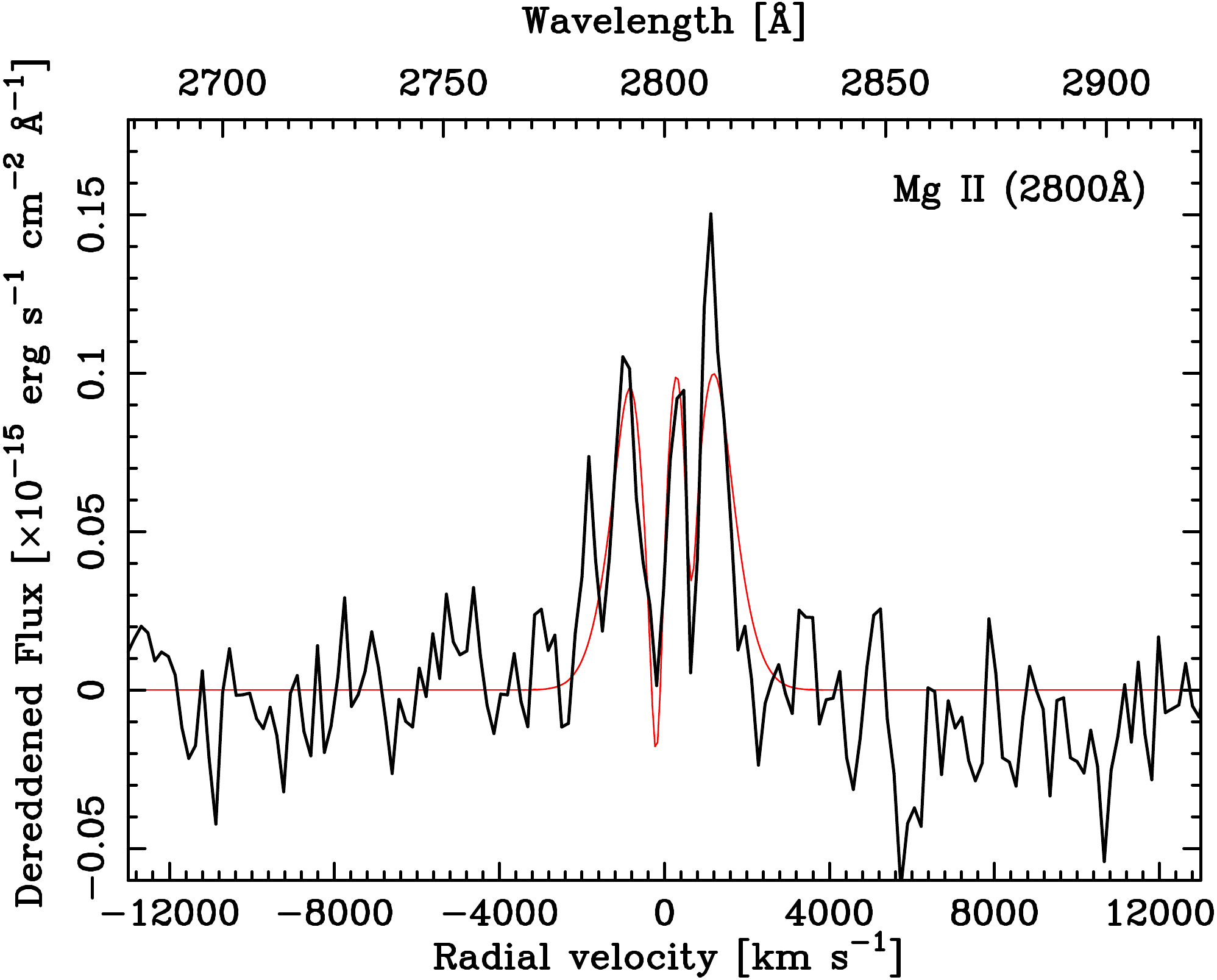}
\caption{Spectral line features for N\,{\sc v} (1240\,\AA; top left), Si\,{\sc iv}\,/\,O\,{\sc iv} (1400\,\AA; top right), N\,{\sc iv}] (1486\,\AA; middle left), C\,{\sc iv} (1550\,\AA; middle right), O\,{\sc iii}] (1640\,\AA; bottom left), and Mg\,{\sc ii} (2800\,\AA; bottom right).  The red lines show the model fit to each feature using the He\,{\sc ii} template (see text for details).  For N\,{\sc v} specifically, the combination of a N\,{\sc v} line (blue) and a proposed O\,{\sc ii} line (magenta) are shown (see Section~\ref{n_v}, c.f.\ Figure~\ref{jetplot}).  For He\,{\sc ii} specifically, the spectrum is shown in blue, and the black line is derived from a moving average of 5 pixels.  Here the vertical line denotes the rest wavelength of the transition.\label{line_zoom}}
\end{figure*}

We will now discuss prominent lines from each detected element in turn, the strongest lines are displayed in more detail in Figure~\ref{line_zoom}.

\subsection{Helium}\label{he_ii}

Emission lines from the H-like He\,{\sc ii} are prominent throughout the FUV, NUV, and optical spectra.  Transitions from the odd $n=13-7$ levels (even levels are coincident with the Balmer series) to the $n=4$ level of He\,{\sc ii} (Pickering series) are all observed in the optical.  The $n=4\rightarrow3$ (4687\,\AA) and $5\rightarrow3$ (3204\,\AA) He\,{\sc ii} lines are among the strongest emission lines in the Keck spectrum.  The $n=3\rightarrow2$ (1640\,\AA) transition is clearly observed in the FUV, the $n=4\rightarrow2$ line is coincident with Lyman\,$\alpha$.  All He\,{\sc ii} lines are observed without P\,Cygni absorptions in all spectra.  As the He\,{\sc ii} lines are singlets and relatively un-blended in all spectra, we use the profile of the strongest line (1640\,\AA) as a guideline template for all other observed emission lines.  

The He\,{\sc ii} (1640\,\AA) emission line is symmetrical within the spectral uncertainties, and double-peaked, like the H\,{\sc i} and He\,{\sc i} lines observed in the optical spectra (\otwok, \othreek, see bottom-left plot in Figure~\ref{line_zoom}).  The central `dip' of this profile is at 1640.3\,\AA, consistent (within the spectral resolution) with the rest wavelength of 1640.5\,\AA, confirming the wavelength calibration of the spectrum (the gray horizontal line in the He\,{\sc ii} plot in Figure~\ref{line_zoom} denotes the rest wavelength)The FWHM velocity of this He\,{\sc ii} line (measured by fitting a single Gaussian to the profile) is $2400\pm200$\,km\,s$^{-1}$. This velocity is consistent with the H$\alpha$ line ($v\simeq2400$\,km\,s$^{-1}$; \othreek), at a similar time after the 2014 and 2015 eruptions.  The integrated flux and FWHM velocity of all emission lines discussed in detail are reported in Table~\ref{spec_lines}.

There is evidence for the He\,{\sc ii} (2512\,\AA; $n=7\rightarrow3$) line in the NUV.  This line can also be modeled using the He\,{\sc ii} (1640\,\AA) line profile, but there may be evidence for line narrowing, the FWHM is $1800\pm300$\,km\,s$^{-1}$. Given the systematic line-narrowing with time as reported by \otwok, \othreek, and \ponek, proposed to be due to the ejecta interacting with circumbinary material, such a `deceleration' between the FUV and NUV/Keck spectra is expected.  There is no significant detection of the He\,{\sc ii} (2734\,\AA; $n=6\rightarrow3$) line.

The double-peaked He\,{\sc ii} (4686\,\AA) and H\,{\sc i} lines in the Keck spectrum exhibit asymmetric peaks, as noted by \othreek\ for the early-time H\,{\sc i} lines. The He\,{\sc ii} (4686\,\AA) line detected during the 2014 and 2015 eruptions was blended with the nearby N\,{\sc iii} (4640\,\AA) line and a direct comparison with the Keck 2013 profile is not possible.  

\subsection{Magnesium}

The broad NUV feature at $\sim$2800\,\AA\ is the Mg\,{\sc ii} (2796/2803\,\AA) resonant doublet (Figure~\ref{line_zoom}, bottom right).  A P\,Cygni profile, as seen for the FUV resonance lines (see Sections\,\ref{n_v}--\ref{si_iv}), is not present, but we note that the S/N is much lower and that the NUV epoch is $\sim1.5$\,days after the FUV observation.  The noisy structure of this feature may show four emission peaks and three narrow absorption dips, with the red most two returning almost back to the continuum level.   

The Mg\,{\sc ii} profile can be well represented by a pair of double-peaked He\,{\sc ii} (1640\,\AA) template emission lines, at 2796/2803\,\AA\ with a fixed flux ratio of 13:12 -- the Mg\,{\sc ii} resonant doublet, and narrower absorption lines also at 2796 and 2803\,\AA.  The fit formally doesn't require additional emission or absorption lines. By inspection of the two separate NUV visits, an apparent third feature at 2786\,\AA\ is only present in the earlier visit and may be related to a velocity difference in the Mg\,{\sc ii} profile between the two epochs.  The equivalent width of all measured absorption lines are reported in Table~\ref{table:ism}.  Such deep absorption features are seen in many of the line profiles, typically corresponding to the wavelengths of known interstellar absorption lines.  However, the low spectral resolution necessitated by these observations prohibits us from unambiguously assigning their origin to either ejecta geometry or interstellar absorption. 

\subsection{Nitrogen}\label{n_iv}

The line at 1486\,\AA\ (Figure~\ref{line_zoom}, middle left) is likely to be due to N\,{\sc iv}] (1486\,\AA).  The FWHM of this profile is $2600\pm200$\,km\,s$^{-1}$, but it doesn't show the same double-peaked structure as He\,{\sc ii}, more of a flat top.  However, the line is still well represented by a single template line centered at 1486\,\AA. In the Keck spectrum there is evidence for lines that we propose are also associated with N\,{\sc iv} at 3479, 4606, and 6381\,\AA.

\label{n_v}

The broad and bright feature at 1240\,\AA\ is the commonly observed N\,{\sc v} (1239/1243\,\AA) resonant doublet (Figure~\ref{line_zoom}, top right).  This feature is noticeably broader than the He\,{\sc ii} and N\,{\sc iv} lines, and appears to have a deep P\,Cygni absorption element which is severely blended with the Lyman\,$\alpha$ feature at 1215.7\,\AA.  The red wing of this profile appears particularly broad, compared to the blue wing.  It is punctuated by an absorption line at 1260\,\AA, possibly a blend of interstellar C\,{\sc i} S\,{\sc ii}, or Si\,{\sc ii}.  The N\,{\sc v} feature is notably broader and more asymmetric than that of C\,{\sc iv} (which also has a clear and deep P\,Cygni absorption), therefore we don't believe that the asymmetry of the N\,{\sc v} feature is due solely to the P\,Cygni absorption.  

The N\,{\sc v} P\,Cygni profile may extend as far as 1212\,\AA, giving a terminal velocity of $\lesssim6500$\,km\,s$^{-1}$.  This is consistent with the velocity of the short-lived very high velocity material observed in the visible spectra of the 2015 eruption for $t<1$\,day (\othreek).  \othreek\ proposed that this emission was from optically thin H\,{\sc i} gas moving almost directly toward the observer, possibly in the form of a collimated outflow or even a jet (see Section~\ref{gij}).

The 1240\,\AA\ feature can be modeled by two template lines but with increased FWHM of $4500\pm100$\,km\,s$^{-1}$, and central wavelengths 1238.8\,\AA\ and 1242.8\,\AA\ (the N\,{\sc v} doublet). However, a third emission line at $1261.5\pm0.3$\,\AA, and a, possibly interstellar C\,{\sc i}, absorption line at 1261.5\,\AA\ are required to reproduce the overall gross structure.  In principle such an emission line at $\sim1262$\,\AA\ could be Si\,{\sc ii} (1260\,\AA), O\,{\sc ii} (1261\,\AA), or C\,{\sc i} (1261\,\AA).  The apparent lack of a C\,{\sc ii} emission line at 1334\,\AA\ seems to rule out an identification as C\,{\sc i}, likewise the low ionization energy of Si\,{\sc ii} (16.3\,eV) appears to rule out that species.  This leaves O\,{\sc ii} (1261\,\AA) as seemingly the only option.  In Section~\ref{gij} we discuss an alternative interpretation of this highly asymmetric line profile.  

\subsection{Carbon}

A C\,{\sc ii} (1334\,\AA) emission line is typically seen in novae from both CO \citep{2014A&A...562A..28D} and ONe WDs \citep{1993AJ....106.2408S,2013A&A...553A.123S}.  But this emission line is not seen in the FUV spectrum.  Instead a deep absorption line is seen, possibly interstellar in origin, see Section~\ref{ism_absorb}.  No strong evidence for any C\,{\sc i-iii} emission lines are seen in either the FUV, NUV or Keck spectra.

The feature at 1550\,\AA\ (Figure~\ref{line_zoom}, middle right) is the C\,{\sc iv} (1548/1551\,\AA) resonant doublet.  A double-peaked structure is present in emission and there is a  deep P\,Cygni absorption dip to the blue-side. The two emission peaks do not correspond to the wavelengths of the C\,{\sc iv} lines.  The P\,Cygni absorption extends as far as 1503\,\AA, giving a C\,{\sc iv} terminal velocity of $\lesssim8700$\,km\,s$^{-1}$.

The C\,{\sc iv} profile isn't well reproduced by a single emission line, however it can be well modeled by a pair of template lines centered at 1548.2 and 1550.8\,\AA\ with a fixed flux ratio of 10:9, and an absorption line at 1548\,\AA.  We note that a shift of the C\,{\sc iv} emission lines $0.74\pm0.07$\,\AA\ to the red produces a superior fit.

The presence of a strong (possibly) interstellar C\,{\sc iv} (1548\,\AA) absorption to explain the central `dip' in the overall C\,{\sc iv} profile at 1548\,\AA\ seems unlikely, particularly as a similar absorption at 1551\,\AA\ is not required. It is possible that this dip is simple due to variations in the underlying C\,{\sc iv} emission line profiles (see the discussion in Section~\ref{he_ii}).

\subsection{Silicon}\label{si_iv}

The structure seen around 1400\,\AA\ (Figure~\ref{line_zoom}, top right) may be arbitrated to either the Si\,{\sc iv} resonant doublet, or O\,{\sc iv}] lines.  With many novae first displaying Si\,{\sc iv} which wanes to reveal waxing O\,{\sc iv}] lines.  The profile is three-peaked with a P\,Cygni form.  As the other resonance lines (N\,{\sc v} and C\,{\sc iv}) in the FUV exhibit P\,Cygni profiles, it is probable that there is some contribution from Si\,{\sc iv} here.  The P\,Cygni absorption may extend as far as 1353\,\AA\ giving a terminal velocity of $\lesssim9400$\,km\,s$^{-1}$.

The 1400\,\AA\ profile cannot be well reproduced by solely O\,{\sc iv} template lines; although the red wing can be modeled there is significant excess blue flux.  But the 1400\,\AA\ profile is very well reproduced by {\it solely} the Si\,{\sc iv} (1394/1403\,\AA) resonant doublet with a fixed flux ratio of 15:12.  The fit formally doesn't require any O\,{\sc iv} lines, nor any absorption lines.  Therefore, we conclude that at this epoch there is no significant O\,{\sc iv} emission.

We note that there is no strong evidence for the presence of any O\,{\sc iv} lines at this stage across the FUV, NUV, or Keck spectra.  The ionization energy of O\,{\sc iii} (54.0\,eV) is much greater than that of Si\,{\sc iii} (33.5\,eV).  Therefore, at this stage in the evolution ($t=3.32$\,days), one possibility is that any oxygen is yet to be sufficiently ionized.  However, with the dominant spectral feature being a seemingly optically thick N\,{\sc v} line (ionization energy of N\,{\sc iv} is 77.5\,eV), any oxygen in the ejecta must still be sufficiently shielded from the central radiation field.

\begin{table*}
\caption{Observed and non-observed emission and P\,Cygni lines in the {\it HST} spectra of the 2015 eruption of \novak.\label{spec_lines}}
\begin{center}
\begin{tabular}{llllllll}
\hline
\hline
Line & $\lambda_0$ & FWHM & Terminal Velocity\tablenotemark{a} & Observed Flux & Unabsorbed flux\tablenotemark{b} & \multicolumn{2}{c}{Ionization energies\tablenotemark{c,1}}\\
 & [\AA] & [km\,s$^{-1}$] & [km\,s$^{-1}$] & \multicolumn{2}{c}{$[\times10^{-15}$\,erg\,s$^{-1}$\,cm$^{-2}$]} & \multicolumn{2}{c}{[eV]}\\
\hline
N\,{\sc v}\tablenotemark{d}   & 1239/1243 & $4500\pm100$ & $\lesssim6500$ & $52.2\pm1.0$& $52.5\pm1.0$ & \phn77.5 & \phn97.9\\
O\,{\sc ii}\tablenotemark{d}   & 1261 & $4500\pm100$ & \nodata & $22.3\pm1.3$ & $24.1\pm1.3$ & \phn13.6 & \phn35.1\\
O\,{\sc v} & 1381 & \nodata & \nodata & $<1.2$ & \nodata & \phn77.4 & 138.1 \\
Si\,{\sc iv}  & 1394/1403 & $4800\pm100$ & $\lesssim9400$ & $20.0\pm0.6$ & \nodata & \phn33.5 & \phn45.1\\
O\,{\sc iv}]  & 1401/1407 & \nodata & \nodata & $<1.0$ & \nodata & \phn54.9 & \phn77.4\\
N\,{\sc iv}]  & 1486 & $2400\pm200$ & \nodata & $5.28\pm0.72$ & \nodata & \phn47.4 & \phn77.5\\
C\,{\sc iv}   & 1548/1551 & $3000\pm250$ & $\lesssim8700$ & $20.4\pm0.7$ & $22.6\pm0.7$ & \phn47.9 & \phn64.5\\
Ne\,{\sc v}  & 1575 & \nodata & \nodata & $<0.9$ & \nodata & \phn97.2 & 126.2\\
Ne\,{\sc iv} & 1602 & \nodata & \nodata & $<0.8$ & \nodata & \phn63.4 & \phn97.2\\
He\,{\sc ii}  & 1640 & $2400\pm200$ & \nodata & $3.88\pm0.53$ & \nodata & \phn24.6 & \phn54.4\\
O\,{\sc iii}]  & 1661 & $2400\pm200$ & \nodata & $3.41\pm0.48$ & \nodata & \phn35.1 & \phn54.9\\
O\,{\sc iii}]  & 1667 & $2400\pm200$ & \nodata & $2.00\pm0.51$ & \nodata & \phn35.1 & \phn54.9\\
\hline
O\,{\sc iii}] & 1667 & $1200\pm100$ & \nodata & $1.40\pm0.46$ & \nodata & \phn35.1 & \phn54.9\\
He\,{\sc ii} & 2512 & $1800\pm300$ & \nodata & $0.75\pm0.20$ & \nodata & \phn24.6 & \phn54.4\\
He\,{\sc ii} & 2734 & \nodata & \nodata & $<0.43$ & \nodata & \phn24.6 & \phn54.4\\
Mg\,{\sc ii} & 2796/2803 & $1800\pm300$ & \nodata & $2.44\pm0.55$ & $3.89\pm0.59$ & \phn\phn7.6 & \phn15.0\\
\hline
\end{tabular}
\end{center}
\tablenotetext{a}{Terminal velocity quoted where there is evidence for a P\,Cygni absorption profile.}
\tablenotetext{b}{Unabsorbed flux is estimated under the assumption that any strong absorption troughs affecting the line profiles are caused by interstellar lines, see Table~\ref{table:ism} for estimates of the equivalent widths of these absorption features.}
\tablenotetext{c}{Here we note the energy required to raise the element to the quoted oxidation state and to the subsequent oxidation state, respectively.}
\tablenotetext{d}{Here we work under the assumption that the N\,{\sc v} profile is a combination of N\,{\sc v} and O\,{\sc ii} (see Section~\ref{n_v}), the total flux of the N\,{\sc v} feature can be found by combining the flux of these two components.  O\,{\sc ii} is one proposed explanation for the asymmetry of the N\,{\sc v} profile, see Section~\ref{gij} for further discussion.}
\tablerefs{(1)~\citet{NIST_ASD}
}
\end{table*}

\begin{table}
\caption{Equivalent width of potential interstellar absorption features in the \novak\ spectra.\label{table:ism}}
\begin{center}
\begin{tabular}{lll}
\hline
\hline
Line & $\lambda_0$ & Equivalent\\
& [\AA] & width [\AA]\\
\hline
C\,{\sc i}$^{\dag,\ddag}$ & 1260 & $1.49\pm0.39$\\
O\,{\sc i} & 1304 & $5.04\pm0.87$\\
C\,{\sc ii} & 1335 & $7.20\pm0.79$\\
C\,{\sc iv}$^{\dag,\S}$ & 1547 & $1.63\pm0.29$\\
Mg\,{\sc ii}$^{\dag}$ & 2796 & $3.20\pm0.58$\\
Mg\,{\sc ii}$^{\dag}$ & 2803 & $2.09\pm0.48$\\
\hline
\end{tabular}
\end{center}
\tablenotetext{\dag}{Equivalent widths computed using additional absorption line require to model the overall emission line profile of a separate line.}
\tablenotetext{\ddag}{This absorption feature may be a blend of interstellar C\,{\sc i}, Si\,{\sc ii}, and S\,{\sc ii}.}
\tablenotetext{\S}{This apparent absorption feature may simply be related to the blending of the line profiles of the C\,{\sc iv} resonant doublet.}
\end{table}

\subsection{Oxygen}

As already mentioned, no O\,{\sc i}, or indeed [O\,{\sc i}], emission lines have yet been observed at any time in spectra of \novak.   O\,{\sc i} (1304\,\AA) emission lines are often observed in the FUV spectra of CO and ONe novae, but no evidence for such emission is seen in the FUV spectrum of \novak.  An absorption line, seen at 1304\,\AA, may be an interstellar line broadened by the instrumental response.  There is no evidence for the O\,{\sc i} 8446\,\AA\ line in the Keck spectrum.

In Section~\ref{n_v} we introduced an interpretation of the N\,{\sc v} `red wing' being the presence of a strong O\,{\sc ii} (1261\,\AA) emission line.  On this basis, we can also tentatively assign emission lines seen in the Keck spectrum at 4133 and 4157\,\AA\ as O\,{\sc ii} lines.

O\,{\sc iii} emission is observed across the FUV, NUV, and Keck spectra.  The O\,{\sc iii} (1667\,\AA) line is seen in the overlapping region of the FUV and NUV spectra at $t=3.32$\,days and 1.5\,days later, respectively. In the FUV, there is also a strong line at 1661\,\AA, which could be O\,{\sc iii} (1661\,\AA).  The FUV O\,{\sc iii} lines have a similar line profile to the He-line, but this narrows to $1200\pm100$\,km\,s$^{-1}$ in the NUV.  Unfortunately the O\,{\sc iii} (2322\,\AA) line is not visible in the NUV, which when combined with the FUV lines may have granted some insight on the electron temperature and density \citep[see, e.g.,][]{1987ApJ...319..403K}.  In the Keck spectrum, generally strong emission lines, tentatively identified as O\,{\sc iii}, are observed at 3059, 3133, 3265, and 3444\,\AA.  

As discussed in Section~\ref{si_iv}, there is no evidence for O\,{\sc iv} lines in the FUV or NUV.  There is also no evidence in the FUV of the O\,{\sc v} (1371\,\AA) line which is commonly seen in novae from CO WDs.  Throughout the FUV, NUV, and visible spectra, O is not observed to be more than doubly-ionized.

Forbidden emission lines of [O\,{\sc ii}] and [O\,{\sc iii}] are commonly observed in the spectra of novae as they enter the nebular phase.  Although not causally connected, the onset of the nebular phase often coincides with the turn-on of the SSS X-ray emission, which occurred at $t=5.6\pm0.7$\,days (\othreek) for \novak\ in 2015.  Based on the behaviour of the optical light curves of the 2013--2015 eruptions, \othreek\ predicted that the nebular phase may have begun between 4 or 5 days post-eruption.  The NUV and Keck spectra were taken $\sim1$\,day before the SSS turn-on but show no evidence for any forbidden O (or N) lines.

\subsection{Neon: The elephant in the room}

When we are concerned with the ultimate fate of an accreting WD, the underlying composition and the deduced mass of the object are of crucial importance. Following \citet{1985ApJS...58..661I} and \citet{2016JPhCS.665a2008S}, it is widely believed and often claimed that only massive stars can form ONe WDs, and that those WDs are massive (i.e.\ $\>1.2-1.3\,M_\odot$). Thus the presence of strong Ne lines is offered as evidence that a WD is not only composed of O and Ne, but that it must be massive. We note that it may be possible to build an 0.1\,M$_\odot$ ONe envelope on a moderate mass CO WD via rapid mass transfer \citep{1994AJ....107.1542S} in a binary, so the presence of Ne does not prove that the underlying WD is particularly massive. We are unaware of any eclipsing, double-lined spectroscopic binaries in old ONe novae for which an accurate mass estimate of the ONe WD is available.

That aside, the [Ne\,{\sc v}] (1575\,\AA) and [Ne\,{\sc iv}] (1602\,\AA) emission lines have only been observed in nova eruptions from ONe WDs.  Ne is not produced in any large quantities by the CNO nuclear reactions on the WD surface.  Therefore, any Ne overabundance observed in the ejecta must be caused by dredge up from the underlying (ONe) WD due to mixing between the accreted layer and the WD; also see the discussion in Section~\ref{sec:wd}.

It is clear from the FUV spectrum that Ne lines are not detected.  Using the  He\,{\sc ii} template, we can place $3\sigma$ upper limits on the flux of [Ne\,{\sc v}] (1575\,\AA) and [Ne\,{\sc iv}] (1602\,\AA) lines of $<9.49\times10^{-16}$\,erg\,s$^{-1}$\,cm$^{-2}$ and $<7.75\times10^{-16}$\,erg\,s$^{-1}$, respectively.

Novae Pupis 1991 and Cygni 1992 also showed evidence for permitted Ne\,{\sc iii} lines at 2040 and 2150\,\AA\ \citep{1993AJ....106.2408S}, but again no signs of such lines are seen in the \novak\ NUV spectrum.  A relatively weak emission line is seen in the Keck spectrum at 3343\,\AA, which {\it could} correspond to a [Ne\,{\sc iii}] line but also O\,{\sc iii} lines.  We note that there is no evidence for the corresponding [Ne\,{\sc iii}] (3869\,\AA) line in that spectrum.  The Keck spectrum also shows no evidence for the [Ne\,{\sc v}] (3346, 3426\,\AA) doublet.

\subsection{Interstellar absorption lines}\label{ism_absorb}

As discussed in the previous sub-sections, a number of absorption lines are seen in the FUV and NUV spectra of \novak.  In many cases the central wavelengths of these absorptions lines match those of expected interstellar absorption lines.  The low spectral resolution employed in these observations is restrictive when assigning a cause to these absorption features, arising either from ejecta or interstellar absorption.  However, the consistency of the appearance (or not) of absorption features within the broad line profiles with the location of expected interstellar lines is telling.  In Table~\ref{table:ism} we compile the proposed interstellar absorption lines observed in the UV spectra of \novak\ and record their estimated equivalent widths.

\subsection{Comparison to LMCN\,1968-12a}

\begin{figure}
\includegraphics[width=\columnwidth]{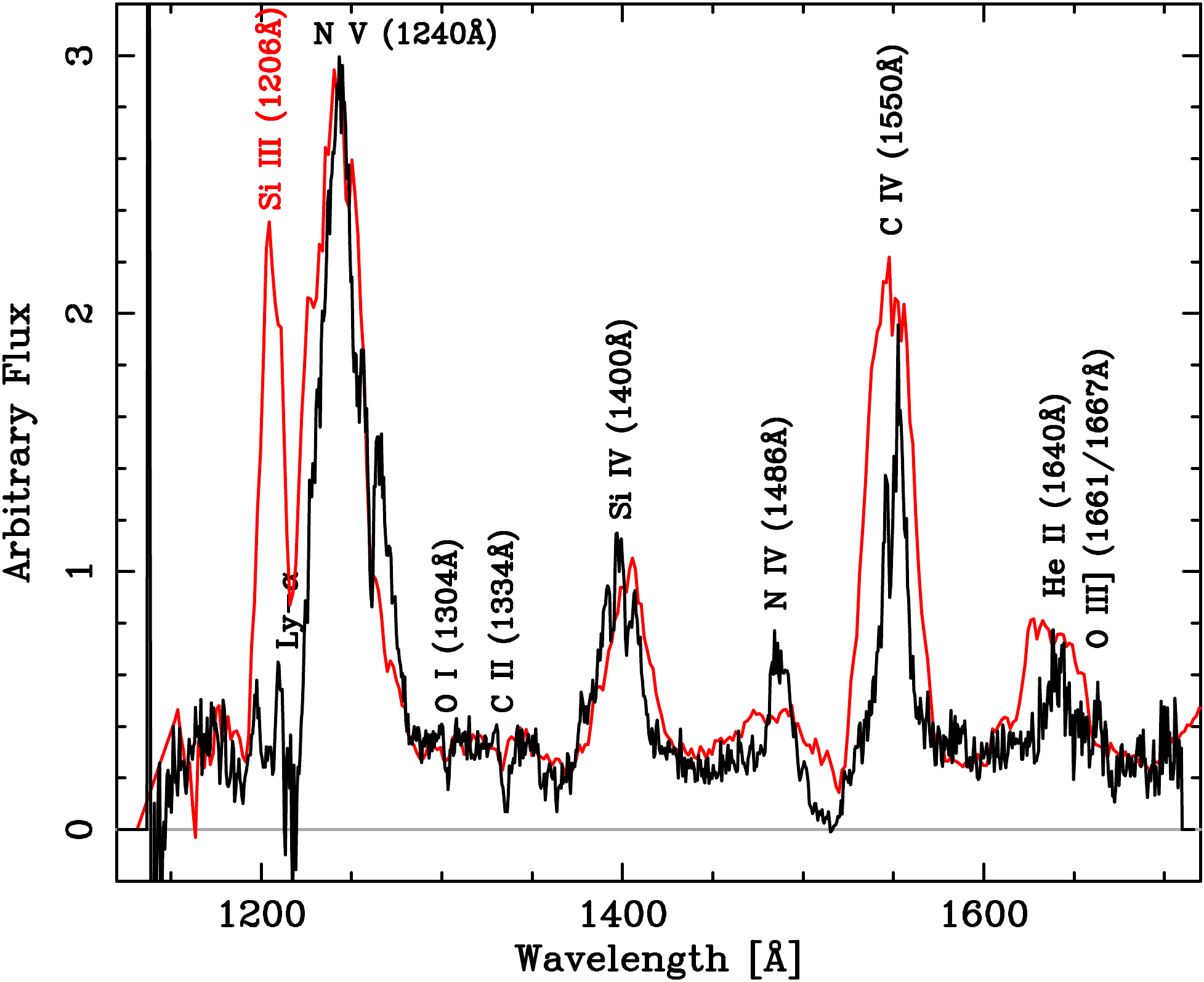}
\caption{A direct comparison between the FUV spectrum of \novak\ (black; cf.\ Figure~\ref{main_spec}) and the IUE spectrum of LMCN\,1968-12a (red) obtained 2\,days post-eruption \citep[see][]{1991ApJ...370..193S}. In general both spectra are similar, but note the differences in line widths and the Si\,{\sc iii} (1206\,\AA) and N\,{\sc iv}] (1486\,\AA) lines.\label{1968}}
\end{figure}

The FUV spectrum of the 2015 eruption of \novak\ shows similarity to the early-time FUV spectrum of the 1990 eruption of the recurrent LMCN\,1968-12a \citep{1991ApJ...370..193S,KuinPaper}.  In Figure~\ref{1968} we directly compare the FUV spectrum of \novak\ to the International Ultraviolet Explorer (IUE) spectrum of LMCN\,1968-12a obtained two days after the 1990 eruption.  The IUE spectra were retrieved from the IUE Final Archive\footnote{\url{https://archive.stsci.edu/iue}}.

In general, these spectra are similar and are dominated by the resonance emission lines.  The C\,{\sc iv} line in LMCN\,1968-12a is noticeably broader, but the P\,Cygni absorptions to the blue of the Si\,{\sc iv} feature and C\,{\sc iv} are less deep and extend to lower velocities.  We also note the lack of an O\,{\sc i} (1304) emission line in both novae, although [O\,{\sc i}] lines were later observed in the visible spectra of LMCN\,1968-12a  \citep{1991ApJ...370..193S}.  \novak\ exhibits a much stronger N\,{\sc iv}] line, and there may be some indication (based on the apparent redward-shift of the profile) that the Si\,{\sc iv} feature in LMCN\,1968-12a may have some contribution from O\,{\sc iv}] (1401/1407\,\AA).

One of the major differences is around the N\,{\sc v} feature.  In the 1990 eruption of LMCN\,1968-12a there is a strong emission feature to the blue which \citet{1991ApJ...370..193S} identified as possibly Si\,{\sc iii} (1206\,\AA).  The central peak of the N\,{\sc v} profile appears slightly broader in LMCN\,1968-12a, however the extended red-wing as seen in \novak\ is not observed (see Section~\ref{gij}).  We note that LMC\,1968-12a did not go on to develop Ne lines in its spectrum (see also Section~\ref{sec:wd}).

\subsection{Continuum emission}

Continuum emission from \novak\ is clearly detected in the FUV, NUV, and Keck spectra.  The bottom panel in Figure~\ref{main_spec} shows a combination of these three spectra, plotted on a log--log scale. We assume that a combination of the FUV and NUV spectra, taken $\sim$1.5\,days apart, is suitable for discussion purposes at least.  To match the NUV and Keck spectra we have artificially increased the Keck flux by 50\%.  Such a flux discrepancy may be due to a number of reasons including, spectra taken at different stages of the eruption, slit losses, differences in background subtraction, etc., and we do not deem it to be significant.

The \novak\ continuum at this epoch is well represented by a power-law across most of the wavelength range (1000--10000\,\AA).  There is clear evidence for the Balmer discontinuity (the Balmer limit, 3645\,\AA, is indicated by the gray-dashed vertical line), and tentative evidence for the Paschen discontinuity (8204\,\AA).  The solid red line in the bottom of Figure~\ref{main_spec} shows a power-law, with the break to account for the Balmer jump.  The continuum emission over the wavelength range $1600\lesssim\lambda\lesssim6000$\,\AA\ is consistent with the power-law associated with optically {\it thick} free-free emission ($f_\lambda\propto\lambda^{-8/3}$).  Such good agreement increases our confidence in the extinction value derived in Section~\ref{sec:extinction}.  Beyond 6000\,\AA, the continuum is roughly consistent with a power law with index $-2.1$, as would be expected for optically {\it thin} free-free emission.  We note that background subtraction and stellar crowding issues may be becoming problematic at longer wavelengths in the Keck data, so this apparent break at longer wavelengths may be due to source contamination.  However, we conclude that the NUV and optical continuum emission at this stage is dominated by optically {\it thick} free-free, while at longer wavelengths the transition to optically {\it thin} free-free may already be underway.

In the bottom panel of Figure~\ref{main_spec} we also show the broadband spectral energy distributions (SEDs; magenta points; from the LT and {\it Swift}) at a similar epoch (see \othreek).  Due to the presence of an emission line spectrum, the SED over-estimates the continuum.  Based on these data, \othreek\ suggested that the continuum was dominated by optically thick free-free emission; the NUV and Keck spectra confirm this.

There is a clear break to an {\it apparent} flatter continuum as we move to the FUV regime.  We also note that such a phenomenon was reported in the broadband SEDs following the 2015 eruption (\othreek).  With the SSS turn-on imminent, we do not believe that this flattening is evidence fot a turnover in the continuum emission.  There could be a number of explanations for this behavior.  Partly it may be due to the $1.5$\,day offset between the FUV and NUV spectra.  It could also be due to variation in intra-ejecta absorption as the ejecta are still undergoing early-time rapid expansion.  But we propose that it is most probably to be due to extinction effects.  The extinction at FUV wavelengths is particularly sensitive to the composition, distribution, and size of the intervening dust grains.  A small change in the reddening or extinction coefficient can cause a large effect at FUV wavelengths, and cannot be ruled out here.

\section{Discussion}\label{sec:disc}

\subsection{Geometry, Inclination, and Jets?}\label{gij}

All the spectral evidence (see \oonek, \otwok, \othreek, \ponek) point to the \novak\ ejecta being highly asymmetric. \othreek\ suggested a geometry similar to that proposed for the $\gamma$-ray nova V959\,Monocerotis \citep{2014Natur.514..339C}.  That is a dense equatorial ejecta whose expansion is impinged by the disk, the donor, and circumbinary material, combined with a lower density, higher velocity polar outflow.  

\othreek\ presented two different interpretations of the system inclination based on the visible and X-ray observations.  The visible spectra, which showed short-lived, high-velocity material, were interpreted as evidence for collimated outflows, possibly jets, aligned almost directly towards the observer -- a low inclination system. Whereas a tentative detection of emission lines in the SSS spectrum could be suggestive of a high inclination \citep[see][for a full and detailed discussion of such obscuration effects]{2013A&A...559A..50N}, however such lines can only be verified using high resolution grating spectra.

The double-peaked spectral lines seen throughout the optical spectra (\oonek, \otwok, \othreek, \ponek) are reproduced in the FUV and NUV spectra, and all but one emission line appears symmetrical.  We propose that this double-peaked line profile, as seen for the He\,{\sc ii} lines and used to model the spectral lines, is due to the equatorial component of the ejecta in a low inclination system.  Again the high quiescent NUV disk luminosity reported by \oonek\ and \ponek\ effectively rules out a high inclination system.

The emission portion of the N\,{\sc v} (1240\,\AA) feature in the FUV appears strongly asymmetric.  In Section~\ref{n_v} we presented one possible explanation, a strong high-velocity O\,{\sc ii} (1261\,\AA) emission line.  However, a search of the literature provided no evidence for such a line ever being observed in a nova spectrum; let alone one as strong as the Si\,{\sc iv} (resonant) emission lines.  So here we present an alternative explanation.  In the $t<1$\,d optical spectra of the 2015 eruption, \othreek\ proposed that the high velocity emission seen at either side of the central H$\alpha$ emission peak was due to highly collimated polar outflows directed almost directly toward the observer.  \othreek\ modeled this high velocity emission with a pair of Gaussians of width $2800\pm100$\,km\,s$^{-1}$ offset from the central wavelength by $-4860\pm200$\,km\,s$^{-1}$ and $+5920\pm200$\,km\,s$^{-1}$ (although the red component was blended with He\,{\sc i} (6678\,\AA) emission which probably accounts for the higher apparent velocity). 

\begin{figure}
\includegraphics[width=\columnwidth]{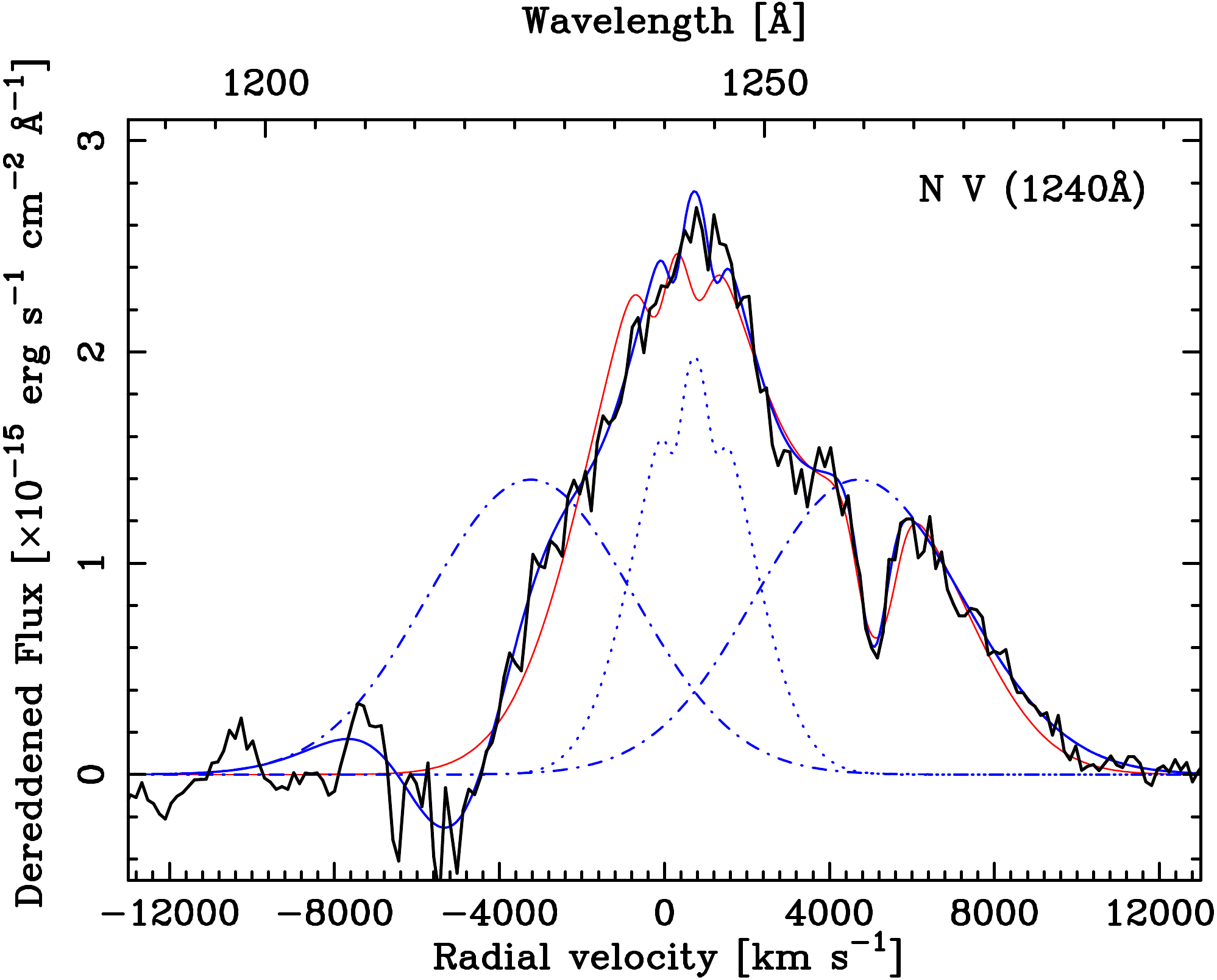}
\caption{Spectral line features for N\,{\sc v} (1240\,\AA).  The red line shows the synthetic N\,{\sc v} + O\,{\sc ii} profile (as detailed in Figure~\ref{line_zoom}.  The solid blue line shows the N\,{\sc v} + optically thick jet profile.  The blue dotted line indicates the equatorial N\,{\sc v} profile (we note the similarity to the C\,{\sc iv} profile), and the dashed-dotted blue line shows the proposed jets (the fore-jet would be heavily self absorbed and affected by Ly$\alpha$).\label{jetplot}}
\end{figure}

The N\,{\sc v} feature can be equally well fit by the N\,{\sc v} doublet, an (ISM) absorption feature at $\sim1260$\,\AA, and a Gaussian emission feature at $+4810\pm250$\,km\,s$^{-1}$.  This is of course remarkably similar to the proposed outflow velocity seen in the optical.  The {\it observed} N\,{\sc v} feature clearly cannot trivially support the presence of another outflow emission feature at $-4810\pm250$\,km\,s$^{-1}$ due to its coincident with the apparent P\,Cygni absorption of N\,{\sc v} and the Lyman\,$\alpha$ feature.  However, as there probably is optically thick N\,{\sc v} present at that epoch, we must consider that the entire N\,{\sc v} feature {\it could} be described by a central peak from equatorial ejecta, close to the plane of the sky, and optically thick (in N\,{\sc v}) outflows in a direction close to the line of sight ($v=\pm4000\pm600$\,km\,s$^{-1}$).  Such a scenario could even partly explain the unidentified emission lines at $\sim$1197 and $\sim$1211\,\AA, see Figure~\ref{jetplot}.  We note that the model including the jet formally provides a marginally better fit (despite the decreased degree of freedom), and reduces or even removes the requirement for a broadening of the model line profile (based on the He\,{\sc ii} profile) of the equatorial emission.

If this is indeed the case, the picture could fit quite well with the presence of such an apparently strong  N\,{\sc v} feature.   N\,{\sc v} is the highest ionization energy species seen in the FUV spectra yet, given the above picture, this material may attain some of the highest velocities in the ejecta.  There is no evidence for the lower ionization O\,{\sc iv} species, yet N\,{\sc iv} and O\,{\sc iii} are seen in abundance and at lower velocities.  The implication must be that we are observing N\,{\sc v} emission from a different region of the ejecta.  Could this simply be explained as material in the fast expanding, low density, outflows seeing a harder radiation field from the WD than material in the denser equatorial regions?  Indeed the equatorial ejecta would be partly shielded by the proposed surviving disk.

Such a discussion wouldn't be complete without comparison to the derived geometries of other novae.  The Galactic RN T\,Pyxidis and the suspected recurrent KT\,Eridani both exhibited similar line profiles to \novak\ in the visible (see \citealt{2014AJ....147..107S} and \citealt{2013MNRAS.433.1991R}, respectively).  Subsequent modeling of these ejecta by \citet{2013A&A...549A.140S} and \citet{2013MNRAS.433.1991R}, respectively, reported different geometries and inclinations to those proposed for \novak.  However, when studying near-IR spectra of T\,Pyx, \citet{2011A&A...534L..11C} reported a similar geometry -- `a face-on bipolar event' -- to our \novak\ model.  We must note two potentially key differences between the line profiles of \novak\ and those of T\,Pyx and KT\,Eri.  Broad rectangular `shoulders' are seen around doubled peaked Gaussian-like central components in the line profiles of all three novae.  But the velocities of these shoulders (see \othreek) are significantly higher in \novak\ and only appear fleetingly in the visible spectra ($t<1$\,day post-eruption; \othreek), whereas they persist throughout the evolution of T\,Pyx and  KT\,Eri.

\subsection{The white dwarf}\label{sec:wd}

\citet{2013A&A...553A.123S} presented `template' FUV spectra of nova eruptions from ONe WDs.  It is clear that the \novak\ FUV spectrum $t=3.32$\,days does not match those templates.  Primarily because there are no Ne lines observed in that spectrum.

In addition, the FUV spectrum of \novak\ hints at being under abundant in oxygen.  The O\,{\sc i} (1305\,\AA) line is simply not observed (only an absorption line is seen, possibly originating from the ISM), no O\,{\sc iv} is required to explain the feature at 1400\,\AA\ (which is Si\,{\sc iv}), and the O\,{\sc iii} line at 1667\,\AA\ appears significantly suppressed compared with the C\,{\sc iv} (1550\,\AA) feature -- spectra of ONe novae often show these features with similar strength.

The FUV spectrum of \novak\ appears significantly under abundant in both oxygen and neon, when compared to the ONe `template' spectra.  The \novak\ spectra are dominated by H, He, and N.

But can we say with certainty that the WD in the \novak\ system is CO in nature?  No we cannot.  The [Ne\,{\sc iv}] and [Ne\,{\sc v}] lines that are the hallmark of ONe novae require high ionization energies and, as forbidden transitions, are only present below a critical density in the nova ejecta.  No nebular lines of any species are observed up to day $\sim5$, and the only species observed with similar ionization levels to [Ne\,{\sc iv}] is N\,{\sc v}.  We have already noted our thoughts about that line, and again note that those are resonance lines.  However, we only have one FUV epoch.  To absolutely rule out the presence of Ne in the ejecta we require a time series of FUV spectra further into the decline.  

We must also pose the question, can we ever know the composition of the WD in such a RN?  That is, without awaiting the final outcome once the Chandrasekhar mass is  exceeded.  For WD material to reach the ejecta that material must first become sufficiently mixed with the accreted envelope -- the nuclear burning region.  The presence of Ne in the ejecta of a number of CNe \citep{1985MNRAS.212..753W,1987MNRAS.228..329S}, plus theoretical studies \citep[see, e.g.,][]{2016A&A...595A..28C}, supports that this is indeed possible.  However, if the mass of the WD is growing with time the situation may differ.  

Each RN eruption must deposit some of its He `ashes' back onto the surface of the WD, each subsequent RN eruption deposits additional He.  Over time a large layer of He accumulates which may act to `shield' the accreted H-rich layer from the WD material.  For significant mixing of WD material into the accreted layer to occur, the mixing timescale must be at most similar to the recurrence time-scale.  But as a given WD grows toward the Chandrasekhar mass its recurrence period decreases eventually (potentially) becoming shorter than the mixing timescale.

If we return to the growing He layer beneath the accreted envelope.  Predictions show that eventually this layer will grow dense and hot enough to initiate He-burning and a He-nova will ensue \citep{1993A&A...269..291J,1998ApJ...496..376C}.  With models suggesting that the He-nova recurrence period is $\sim1000$ times longer than that of the parent system's H-novae, it seems highly probable that WD material will always have time to mix with the He-layer.  We predict that the ejecta from He-novae will contain the abundance signature of the underlying WD, but these are rare events -- and may not eject significant quantities of material \citep{2016ApJ...819..168H}.

In effect, mixing of WD material into the ejecta of short period RNe may be strongly suppressed.  As such, so is the possibility of one being able to determine the composition of an accreting massive WD.  We note that there is {\it no evidence} of Ne in any of the Galactic RNe, above Solar abundances \citep[see, e.g.,][]{2013A&A...556C...2M}. 

However, if confirmed, the lack of Ne in the \novak\ ejecta does tell us one important thing.  That WDs in accreting systems can indeed increase their mass.  In \novak, either we have a near-Chandrasekhar CO WD (i.e.\ grown in mass), or the system contains an ONe WD that has accumulated a large enough He layer (i.e.\ grown in mass) to shield the internal WD from the accreted layer and nuclear burning.

\section{Summary \& Conclusions}\label{sec:conc}

In this paper we have analyzed unique {\it Hubble Space Telescope} spectroscopic observations of the 2015 eruption of \novak.  In this analysis, we have also utilized an archive Keck spectrum of the 2013 eruption of \novak.  Here we summarize our conclusions:

\begin{enumerate}
\item A series of {\it HST} STIS NUV spectra of the 2015 eruption aided constraining the reddening toward \novak\ to $E_{B-V}=0.10\pm0.03$ -- consistent with all the line-of-sight extinction being internal to the Milky Way.
\item A single {\it HST} STIS FUV spectrum showed no indication of any neon ([N\,{\sc iv}] 1602\,\AA\ or [N\,{\sc v}] 1575\,\AA) in the ejecta of \novak\ at that time ($t=3.32$\,d post-eruption).
\item The FUV spectrum was broadly consistent with the post Fe-curtain spectrum of a CO WD.  But we note the absence of emission lines of O\,{\sc i} (1304\,\AA), C\,{\sc ii} (1334\,\AA), and O\,{\sc v} (1371\,\AA).
\item The FUV resonance lines (N\,{\sc v} 1240\,\AA, Si\,{\sc iv} 1400\,\AA, and C\,{\sc iv} 1550\,\AA) all exhibited strong P\,Cygni absorption profiles with mean terminal velocity $\lesssim8200$\,km\,s$^{-1}$.  Other emission lines in the FUV had typically FWHM velocities of 2400\,km\,s$^{-1}$, the emission portion of the resonance lines were all broader. 
\item The NUV spectrum was dominated by the Mg\,{\sc ii} resonance line, but was otherwise somewhat devoid of lines with only a He\,{\sc ii} and O\,{\sc iii} line visible.  There was evidence for significant line narrowing in the $\sim1.5$\,d between the FUV and NUV spectra.
\item The NUV spectrum and the Keck 2013 spectrum were taken at similar epochs, post-eruption.  When combined, the dereddened continuum was consistent with the expected slope of optically {\it thick} free-free emission. 
\end{enumerate}

The primary aim of these {\it HST} observations was to determine the WD composition -- and hence the ultimate fate of the system.  The lack of detected neon, may be indicative of a CO WD, simply looking too early in the eruption, or due to shielding of the WD material from the accreted layer (by a growing He layer).

Although these observations have addressed a number of significant questions about \novak\, they have opened up as many new questions as they have answered.  Most of these questions can only be solved by further observations, particularly time-resolved UV spectroscopy of the eruptions, and further more detailed modeling of this extreme system.

\begin{acknowledgements}

The authors would like to offer their thanks to the anonymous referee of the original manuscript for their constructive comments and suggestions. 

Based on observations made with the NASA/ESA Hubble Space Telescope, obtained from the Data Archive at the Space Telescope Science Institute, which is operated by the Association of Universities for Research in Astronomy, Inc., under NASA contract NAS 5-26555. These observations are associated with program \#14125.

The W.\ M.\ Keck Observatory is operated as a scientific partnership among the California Institute of Technology, the University of California and the National Aeronautics and Space Administration. The Observatory was made possible by the generous financial support of the W. M. Keck Foundation.  The authors wish to recognize the significant cultural role that the summit of Maunakea has always had within the indigenous Hawai'ian community.  We are most fortunate to have the opportunity to conduct observations from this mountain.

This research has made use of the Keck Observatory Archive (KOA), which is operated by the W.\,M.\ Keck Observatory and the NASA Exoplanet Science Institute (NExScI), under contract with the National Aeronautics and Space Administration.  The authors also acknowledge Shri Kulkarni as the PI of Keck program ID C254LA.  We also acknowledge the Keck observer Yi Cao.

The authors would like to thank the {\it HST} staff for their heroic efforts scheduling our early-time, and disruptive, spectroscopic observations.  Particular thanks are given to Charles R.\ Proffitt, the STScI Contact Scientist for programme \#14125, for his support with the STIS observations.

MJD would like to personally thank Julianne Dalcanton for guidance with the PHAT extinction studies, Laura Darnley (n{\'e}e Wisson) for reading chemistry(!), Toby Moore for useful discussions about interstellar extinction, and Steve Shore and Iain Steele for their spectroscopic insight.

PG wishes to thank William (Bill) P.\ Blair for his kind hospitality in the Rowland Department of Physics \& Astronomy at the Johns Hopkins University.

MH acknowledges the support of the Spanish Ministry of Economy and Competitiveness (MINECO) under the grant FDPI-2013-16933 as well as the support of the Generalitat de Catalunya/CERCA programme. 

NPMK acknowledges funding from the UK Space Agency.

SCW acknowledges a visiting research fellowship at Liverpool John Moores University. 

KH was supported by the project RVO:67985815.

VARMR acknowledges partial financial support from the Radboud Excellence Initiative, from Funda\c{c}\~ao para a Ci\^encia e a Tecnologia (FCT) in the form of an exploratory project of reference IF/00498/2015, from Center for Research \& Development in Mathematics and Applications (CIDMA) strategic project UID/MAT/04106/2013 and from Enabling Green E-science for the Square Kilometer Array (ENGAGE SKA) ROTEIRO/0041/2013/022217.

This work has been supported in part by NSF grant AST-1009566.  Support for program \#14125 was provided by NASA through a grant from the Space Telescope Science Institute, which is operated by the Association of Universities for Research in Astronomy, Inc., under NASA contract NAS 5-26555.
\end{acknowledgements}

\facilities{HST (STIS), Keck:I (LRIS)}

\software{gnuplot (v5.0.6), IRAF \citep[v2.16.1,][]{1993ASPC...52..173T}, PGPLOT (v5.2)}

\bibliographystyle{aasjournal} 
\bibliography{refs} 

\end{document}